\documentclass[aps,prd,eqsecnum,showpacs,amsmath,nofootinbib,superscriptaddress,twocolumn,
floats,preprintnumbers]{revtex4-1}
\usepackage[dvips]{color,graphicx,epsfig}
\usepackage{amsfonts,amssymb,theorem,mathrsfs,times}
\usepackage{epsfig}
\usepackage{latexsym, amssymb}

\usepackage{color}
\usepackage{bbold}
\usepackage{graphicx}
\usepackage{dcolumn}
\usepackage{bm}
\usepackage[normalem]{ulem}

\begin{document}
\def\Z#1{_{\lower2pt\hbox{$\scriptstyle#1$}}}

\newcommand{\be}{\begin{equation}}
\newcommand{\ee}{\end{equation}}
\newcommand{\bea}{\begin{eqnarray}}
\newcommand{\eea}{\end{eqnarray}}
\newcommand{\nn}{\nonumber}

\renewcommand{\thefootnote}{\fnsymbol{footnote}}
\long\def\@makefntext#1{\parindent 0cm\noindent \hbox to
1em{\hss$^{\@thefnmark}$}#1}
\def\Z#1{_{\lower2pt\hbox{$\scriptstyle#1$}}}


\title{Natural braneworld inflation in light of recent results from Planck and BICEP2}

\preprint{CERN-PH-TH-2014-189}

\author{Ishwaree P. Neupane}

\affiliation{Theory Division, CERN, CH-1211 Geneva 23,
Switzerland}

\affiliation{Department of Physics and Astronomy, University of
Canterbury, Private Bag 4800, Christchurch 8041, New Zealand}

\affiliation{Centre for Cosmology and Particle Physics, Tribhuvan
University, Kirtipur, Kathmandu 44618, Nepal}

\begin{abstract}

In this paper we report on a major theoretical observation in
cosmology. We present a concrete cosmological model for which
inflation has natural beginning and natural ending. Inflation is
driven by a cosine-form potential, $V(\phi)= \Lambda^4
\left(1-\cos(\phi/f)\right)$, which begins at $\phi \lesssim \pi
f$ and ends at $\phi =\phi_{\rm end} \lesssim 5 f/3$. The distance
traversed by the inflaton field $\phi$ is sub-Planckian. The
Gauss-Bonnet term ${\cal R}^2$ arising as leading curvature
corrections in the action $S = \int d^5{x} \sqrt{-g\Z{5}} M^3
\left(- 6\lambda M^2 + R + \alpha M^{-2} {\cal R}^2\right)+ \int
d^{4}x \sqrt{-g\Z{4}} \left(\dot{\phi}^2/2 - V(\phi)- \sigma
+{\cal L}_{\text matter}\right)$ (where $\alpha$ and $\lambda$ are
constants and $M$ is the five-dimensional Planck mass) plays a key
role to terminate inflation. The model generates appropriate
tensor-to-scalar ratio $r$ and inflationary perturbations that are
consistent with \textit{Planck} and BICEP2 data. For example, for
$N_*= 50-60$ and $n_s\sim 0.960\pm 0.005$, the model predicts that
$M\sim 5.64\times 10^{16}\,{\rm GeV}$ and $r\sim (0.14-0.21)$
[$N_*$ is the number of e-folds of inflation and $n_s$ ($n_{t}$)
is the scalar (tensor) spectrum spectral index]. The ratio
$-n_t/r$ is (13\% -- 24\%) less than its value in 4D Einstein
gravity, $-n_t/r=1/8$. The upper bound on the energy scale of
inflation $V^{1/4}=2.37\times 10^{16}\,{\rm GeV}$ ($r<0.27$)
implies that $(-\lambda \alpha) \gtrsim 75 \times 10^{-5}$ and
$\Lambda<2.17\times 10^{16}\,{\rm GeV}$, which thereby rule out
the case $\alpha=0$ (Randall-Sundrum model). The true nature of
gravity is holographic as implied by the braneworld realization of
string and M theory. The model correctly predicts a late-epoch
cosmic acceleration with the dark energy equation of state ${\rm
w}\Z{\text DE}\approx -1$.

\end{abstract}


\maketitle
\section{Introduction}

Cosmic inflation~\cite{Guth1980,Linde1981} -- the hypothesis that
the Universe underwent a rapid exponential expansion in a brief
period following the big bang -- is a theoretically attractive
paradigm for explaining many problems of standard big-bang
cosmology, including why the Universe has the structure we see
today~\cite{Mukhanov:1981,Mukhanov:1990me} and why it is so big.
It could also solve outstanding puzzles of standard big-bang
cosmology, such as, why the Universe is, to a very good
approximation, flat and isotropic on largest scales.

\smallskip

To get a successful inflationary model that respects various
observational constraints from the {\it Wilkinson Microwave
Anisotropy Probe} (WMAP)~\cite{WMAP}, {Planck}~\cite{Planck2013}
and {\it Background Imaging of Cosmic Extragalactic Polarization}
(BICEP2)~\cite{BICEP2} and other experiments, namely, those
related to the cosmic microwave background (CMB) observations, it
is necessary to obtain an inflationary potential $V(\phi)$ having
the height $V^{1/4}$ much smaller than its width $\Delta\phi$ (the
distance traversed by the $\phi$-field during inflation).
Moreover, observational results from {Planck} ~\cite{Planck2013}
and BICEP2~\cite{BICEP2} lead to an upper bound on the energy
scale of inflation, $V_*^{1/4}= 1.94\times 10^{16}~{\rm GeV}
(r_*/0.12)$, where $r_*$ is the (maximum) ratio of
tensor-to-scalar fluctuations of the primordial power spectra,
while ideas based on fundamental theories of gravity, such as,
superstring and
supergravity~\cite{JAdams:1996,Brandenberger:2000,Neupane:2007},
reveal that $\Delta\phi \sim M\Z{P}$ (where $M\Z{P} = 2.43\times
10^{18}~{\rm GeV}$ is the reduced Planck mass). These two very
different mass scales (differing by at least 2 orders of
magnitude) is what is known as the fine-tuning problem in
inflation. The latter usually requires precise couplings in the
theory so as to prevent radiative corrections from bringing the
two mass scales back to the same level. An inflationary model
parametrized by the following cosine-type
potential~\cite{Freese:1990,Adams:1992,JAdams:1996}:
 \be
V(\phi)= \Lambda^4 \left(1 \pm \cos\left({\phi\over
f}\right)\right),\label{cosine-poten}\ee where $\Lambda\sim
m\Z{G}$ is the vacuum expectation value of the grand unified
theory (GUT) Higgs fields, or the symmetry breaking mass scale of
the GUT $\sim (1 - 2) \times 10^{16}~{\rm GeV}$, avoids this
problem mainly because it uses shift symmetries $\phi=\phi \pm
2\pi f$ to generate a flat potential, which is protected from
radiative corrections in a natural way~\cite{Adams:1992}. The
above potential represents a potential of pseudo Nambu-Goldstone
boson of the grand unified theory, which was initially motivated
by its origin in symmetry breaking in an attempt to naturally give
rise to the extremely flat potentials required for inflationary
cosmology, known as the natural inflation
model~\cite{Freese:1990}.

\medskip

In string theory and some super-gravity models, the effective
scale $\Lambda$ is small compared to $f$ due to the exponential
(instanton) suppression factor, such as, $\Lambda= \alpha\Z{0}
\,e^{-\alpha\Z{1}} f$ and $\alpha\Z{1}\gg
\ln\alpha\Z{0}>0$~\cite{Adams:1992,Ellis:2014,Yonekura:2014}. A
cosine-form potential as in (\ref{cosine-poten}) is obtained also
in particle physics models with weakly self-coupled
(pseudo-)scalars, such as, the axion~\cite{Kim:2004} and the extra
component of a gauge field in a 5D theory compactified on a
circle~\cite{Hosotani:1983,Antoniadis:2001,Arkani-Hamed:2003}. In
the limit of exact symmetry (e.g., supersymmetry), $\phi$ is a
flat direction, so some tilt is necessary for cosmic inflation.
This is provided by explicit symmetry breaking terms, which can be
mediated, for example, by gravitational quantum corrections. It is
thus natural to include the leading-order curvature corrections
also in a gravitational Lagrangian.

\medskip

For the sake of convenience we define $\Lambda^4 \equiv V_0$, so
that \be V(\phi)= V\Z{0} \left[1-\cos\left({\phi\over
f}\right)\right].\label{PNGB}\ee We have taken the negative sign
in (\ref{cosine-poten}) so that $\phi=0$ is the true minimum. It
is straightforward to obtain \be V_\phi^2= {V \over f^2} \left( 2
V\Z{0}- V \right), \quad V_{\phi\phi}= {1 \over f^2} \left(
V\Z{0}- V \right).\label{V-deri}\ee For $\phi \ll f$, $V(\phi)$
gives an approximately quadratic potential, $V(\phi)= m^2 \phi^2$
[with $m^2\equiv \Lambda^4/(2f^2)$], which was studied
in~\cite{Ish14a} in the context of braneworld inflation. In this
limit $V_\phi^2=4 V m^2$ and $V_{\phi\phi}=2 m^2$. Here we work in
a general scenario where $\phi$ is unconstrained. Recently,
in~\cite{Kohri:2014}, it was argued that the natural inflation
model first proposed in~\cite{Freese:1990} and the so-called {\it
extranatural} inflation model~\cite{Arkani-Hamed:2003} can have
distinguishing inflationary signatures.

\medskip

The Planck collaboration and some earlier discussions showed that
in Einstein gravity the potential (\ref{PNGB}) leads to results
compatible with Planck data for inflation if $f\gtrsim (15/\pi)
M\Z{P}$ in the large field limit. The assumption that the inflaton
field $\phi$ may take values larger than the Planck scale and/or
it traverses a distance large compared with the Planck mass during
inflation is outside the range of validity of an effective field
theory description, so it is natural to assume that $f\lesssim
M\Z{P}$. In this paper we show that ${\cal R}^2$-type curvature
corrections in a 5D Lagrangian can remove this drawback of the
original natural inflation model, giving a trustworthy model of
inflation.

\medskip

Another important physical quantity that we would need for
studying impacts of the above mentioned potential on the
primordial scalar and tensor fluctuations is the Hubble expansion
parameter. Einstein gravity often fails to explain the high energy
phenomena. Moreover, according to some fundamental theories of
gravity and particle interactions, including superstring theory,
the true nature of gravity is higher dimensional, whereas the
elementary particles, fundamental scalars, and gauge fields of the
standard quantum field theory live within a four-dimensional
(three dimensions of space and one dimension of time) membrane, or
``brane". This idea is consistent with gravity/gauge-theory
correspondence~\cite{Malda:1997,Witten:1998}, which provides so
far the best understanding of string theory in terms of gauge
field theories, such as, the Yang-Mills theory.

\medskip

As the most natural generalization of Einstein gravity in five
dimensions, we consider the following action ~\cite{Ish:2000} \bea
S &=& S_{\text bulk} + S_{\text brane}\nn \\
&=& \int_{\cal M} d^5{x} \sqrt{|g|} M^3 \left(- 6 \lambda M^2+ R +
{\alpha\over M^{2}} {\cal R}^2\right)\nn\\
&{}& + \int_{\partial {\cal M}} d^4{x} \sqrt{|\tilde{g}|}
\left(-\sigma+ {\cal L}_\phi + {\cal L}_{\rm
matter}\right),\label{main-5Daction}\eea where $\alpha$ and
$\lambda$ are constants, $\sigma$ is the brane tension, $M$ is the
five-dimensional Planck mass, $R$ is the Einstein-Hilbert term,
${\cal R}^2 = R^2 - 4 R_{ab} R^{ab}+ R_{abcd} R^{abcd}$ is the
Gauss-Bonnet (GB) density and ${\cal L}_\phi= \dot{\phi}^2/2 -
V(\phi)$ is the scalar Lagrangian. The GB density, which appears
in the low energy effective action of heterotic string theory and
in Calabi-Yau compactifications of M theory, is known to give
solutions that are free of ghosts about flat and other exact
backgrounds, such as a warped spacetime background~\cite{CNW-01}.
The ${\cal R}^2$ terms can arise as the $1/{\cal N}$ corrections
in the large ${\cal N}$ limit of some gauge theories and thus
provide a testing ground to investigate the effects of
higher-curvature terms in the context AdS/CFT
correspondence~\cite{Cho:2002b,Neupane:2002c}. In this paper we
discuss the wider cosmological implications of the theory.

\medskip

The brane action (also known as boundary action) is crucial to
obtain the correct form of Friedman equations in four dimensions.
The matter Lagrangian ${\cal L}_m$ can be ignored at sufficiently
high energy, $V^{1/4}\gtrsim 10^{15}\,{\rm GeV}$. For a
cosine-form potential given above, inflation begins once the
inflaton field $\phi$ is displaced from $\phi=\pi f$, possibly
breaking a fundamental symmetry of the GUT potential. If the bulk
spacetime is negatively curved [anti-de Sitter (AdS)] $\lambda<0$
and the GB coupling $\alpha>0$, then inflation would have a
natural end. Because of this reason the present model may be
viewed as a "doubly natural inflation" scenario.

\medskip

The recent detection of a gravitational wave contribution to the
CMBR anisotropy by BICEP2~\cite{BICEP2} with a relatively large
tensor-to-scalar ratio $r\sim 0.19\, (+0.007 - 0.005)$ may be
viewed as a clear cosmological gravitational wave signature of
inflation~\cite{Easther:2006}. By reanalyzing the BICEP2 results,
the authors of Ref.~\cite{Flauger:2014} argued that BICEP2 data
are consistent with a cosmology with $r=0.2$ and negligible
foregrounds, but also with a cosmology with $r=0$ and a
significant dust polarization signal (see,
e.g.,~\cite{Abazajian:2014,Liu-Sarkar-2014} for some other
implications of BICEP2 results). This ambiguity may be resolved by
future Keck Array observations at 100 GHz and Planck observations
at higher frequencies.

\medskip

One of the motivations for considering the effects of ${\cal R}^2$
terms on inflationary scalar and tensor perturbation amplitudes is
to obtain a relatively large $r$ that is compatible with the
BICEP2 result, namely, $r=0.19^{+0.07}_{-0.05}$ (or
$r=0.16^{+0.06}_{-0.05}$ after subtracting an estimated
foreground). The value of $r$ reported by BICEP2 is larger than
the bounds $r < 0.13$ and $r < 0.11$ reported by WMAP~\cite{WMAP}
and Planck~\cite{Planck2013}. Many authors have considered various
possibilities~\cite{Hamada:2014,Freese:2014,Bonvin:2014,Csaki:2014,Cai:2014}
for the origin of a cosmological gravitational wave signature that
support a value $r> 0.11$. If the B-mode polarization detected by
BICEP2 is due to primordial gravitational waves, then it implies
that inflation was driven by energy densities at the GUT scale
$m\Z{\rm GUT}\sim \Lambda \sim (1-2) \times 10^{16}\,{\rm GeV}$.
The results in this paper support this idea.

\section{Inflationary parameters}

The Hubble expansion parameter in four dimensions is uniquely
given by~\cite{Ish14a,Davis:2002,Lidsey:2003} \be H^2= {M^2
\psi^2\over |\beta|}\big[(1-\beta) \cosh\varphi
-1\big],\label{main-Fried1}\ee where
\be \varphi \equiv {2\over 3} \sinh^{-1}\left[{\rho_\phi+\sigma
\over \psi M^4}{|2\beta|^{1/2}\over
4(1-\beta)^{3/2}}\right],\label{def-varphi} \ee and \be
 \beta\equiv
4\alpha \psi^2 = 1 \pm \left(1+8\lambda \alpha + {8\alpha {\cal
E}\over a^4 M^2}\right)^{1/2},\ee where $a$ is the scale factor of
the physical universe. We will take the negative root which has a
smooth Einstein gravity or Randall-Sundrum limit ($\alpha=
0$)~\cite{RS2}. ${\cal E}$ is a measure of bulk radiation energy,
which is proportional to the mass of a 5D black hole and $\psi$ is
a dimensionless measure of bulk curvature ($\psi>0$ for an
anti--de Sitter bulk and $\psi<0$ for de Sitter bulk; the $\psi=0$
case which corresponds to a flat 5D Minkowski spacetime must be
treated separately). There are three bulk parameters here:
$\alpha$, $\lambda$, and $\beta$. $\alpha$ and $\lambda$ are taken
to be constants, while $\beta$ would vary with the evolution of
the Universe, especially after reheating since ${\cal E}>0$.
During inflation (more specifically, before reheating) $\beta$ is
also a constant since ${\cal E}\approx 0$ (as there is no
radiation energy or at least not in an appreciable amount) and the
scale factor rapidly grows. The choice $\beta<0$ is also possible,
provided that the bulk is de Sitter ($\lambda>0)$, but we will not
study this case here as it does not lead to a graceful exit from
inflation.

\medskip

In the original braneworld proposal~\cite{RS2} the 3-brane tension
$\sigma$ is assumed to be a constant (which is forced upon only in
the static limit where the Hubble expansion parameter is zero). In
an expanding physical universe, the brane tension can be a
function of the four-dimensional scale factor $a(t)$, or the
volume of the Universe. All the results in this paper are valid
even if $\sigma$ is scale dependent ($\sigma=\sigma(a)$), as long
as the condition $\rho_\phi\gg \sigma$ holds during inflation.

\medskip

As is usually the case, the Hubble expansion parameter is linked
to the four-dimensional scalar-matter density $\rho_\phi\equiv
{1\over 2}\dot{\phi}^2 + V(\phi)$. Further, from
Eq.~(\ref{def-varphi}) we can see that $\rho_\phi$ can be defined
in terms of the energy scale $\varphi$, which is dimensionless.
This is a direct manifestation of holography or
gravity/gauge--theory correspondence. The inflaton equation of
motion is \be \ddot{\phi}(t)+ 3 H(t) \dot{\phi}(t) + V_\phi=0,
\label{main-phi-eq}\ee where $V_\phi\equiv dV/d\phi$. Under the
slow-roll approximation $\ddot{\phi} \ll 3 H(t) \dot{\phi}$,
$\rho_\phi \simeq V \gg \sigma$,~\footnote{The brane tension can
be neglected for inflationary calculations; it finds an
interesting role at low energies, such as during a transition from
the decelerating to accelerating phase.} we get \be V \simeq
{4(1-\beta)^{3/2}\over (2\beta)^{1/2}} \psi M^4 \sinh(3\varphi/2).
\ee By substituting this expression in Eq.~(\ref{V-deri}) we
obtain the slow-roll parameters: \bea \epsilon &=& - {\dot{H}\over
H^2} \simeq {dH\over
d\varphi} {d\varphi \over dV} {V_\phi^2\over 3 H^3}\nn \\
&=& {2\beta (1-\beta) V\Z{0}\over 9 M^2\psi^2 f^2} {\sinh\varphi
\tanh(3\varphi/2) (1-X(\varphi))\over
\left[(1-\beta)\cosh\varphi-1\right]^2},\eea 
\bea \eta = {V_{\phi\phi}\over 3 H^2} = {\beta V\Z{0}\over 3  M^2
\psi^2 f^2} {\left(1-2 X(\varphi)\right)\over
\left[(1-\beta)\cosh\varphi-1\right]},\eea where
$$X(\varphi) \equiv (1-\beta)^{3/2} \sinh\left({3\varphi\over 2}\right) \chi, \quad
\chi\equiv {M^4\over \sqrt{2\alpha} V\Z{0}}.$$ Typically, $V\Z{0}
\sim M^4$ and $\alpha \gtrsim 10^{3}$, so $\chi < 0.05$. After a
few e-folds of cosmic inflation, $X \ll 1$ and $V(\phi)$
approximates to the quadratic $m^2\phi^2$ potential. As we
establish below, the model deviates from the GB assisted
$m^2\phi^2$-inflation~\cite{Ish14a}, especially, at higher
energies ($\varphi\gtrsim 1$).

\medskip

We will work under the assumption that $\beta \ll 1$ and inflation
ends at $\varphi=\varphi_e \ll 1$, where subscript `e' refers to
the end of inflation. We will justify these assumptions. Inflation
ends ($\epsilon \ge 1$) at $\varphi=\varphi_e$ when \be {2\beta
(1-\beta) V\Z{0}\over 9 M^2 \psi^2 f^2} \simeq {\varphi_e^2\over
(6-8\beta)}.\label{inflation-end}\ee The second term above can be
expressed also in terms of the number of {\it e}--folds of
inflation, $N\equiv \int H dt$; the number of {\it e}--folds is
well approximated by \bea N \equiv \int_{\varphi_*}^{\varphi_e} H
{dt\over d\phi} {d\phi\over dV} {d V\over d\varphi} d\varphi
\simeq  3 \int_{\varphi_e}^{\varphi_*} {H^2\over V_\phi^2}
\left(dV\over d\varphi\right) d\varphi, \nn \eea where the
equality holds in the slow--roll approximation $\ddot{\phi} \ll 3
H(t) \dot{\phi}$. $\varphi_*$ is the value of $\varphi$ where the
mode $k_*=a_* H_*$ crosses the Hubble radius for the first time
(during inflation). We will assume that $\chi \sinh(3\varphi_*/2)
< 1$. As a good approximation we have \be N_*\simeq {9 M^2 \psi^2
f^2\over 4 \beta V\Z{0}} \left( I(\varphi_*)-
{(2-5\beta)\varphi_e^2\over 12}\right), \label{N*eqn}\ee where the
function $I(\varphi) $ is well approximated by \bea I(\varphi) &=&
\varphi-{2\beta\over 3}
\ln\left(e^\varphi-1\right) +(1-\beta) (\cosh \varphi-1) \nn \\
&{}& \quad + {3-\beta\over 3}\big[\ln 3-
\ln\left(e^{2\varphi}+e^\varphi+1\right)\big] \eea
From Eqs.~(\ref{inflation-end}) and (\ref{N*eqn}) we establish
that 
\be {2\beta (1-\beta) V\Z{0}\over 9 M^2 \psi^2 f^2} \simeq
{\varphi_e^2\over (6-8\beta)} \simeq {I(\varphi_*)(1-\beta)\over
2N_*+1-5\beta}.\label{matching}\ee 
We can drop the term $5\beta$ since the number of {\it e}--folds
required to explain the flatness and horizon problems of the hot
big--bang cosmology is large, $N_*\sim 50-62$, depending upon the
detail of reheating mechanism after the end of inflation, while
$|\beta| \ll 1$. The above matching condition works well for
$\beta \lesssim 10^{-2}$. To a good approximation, \bea \epsilon
&= & {(1-\beta)\,I(\varphi_*) \over 2N_*+1} {\sinh\varphi_*
\tanh(3\varphi_*/ 2) \left(1-X(\varphi_*)\right)\over
\left[(1-\beta)\cosh\varphi_* -1\right]^2},\nn \\ \\
\eta &=& {3 I(\varphi_*)\over 2(2N_*+1)}
{\left(1-2X(\varphi_*)\right)\over (1-\beta)\cosh\varphi_*-1}.
\eea 
In the discussion below we take $\varphi_* < 2.5 $, so that
$X(\varphi_*) < 0.3 $. As shown in Fig.~1, inflation has a natural
exit ($\epsilon>1$) only if $\beta>0$, which means $\lambda<0$.
The Planck+WP (or WMAP 9-yr large angular scale polarization)
constraints imply $\epsilon < 0.01$ and $\eta < 0.008$ at 95%
CL. This result is fully compatible with the present model.

\begin{figure}[ht]
\begin{center}
\epsfig{figure=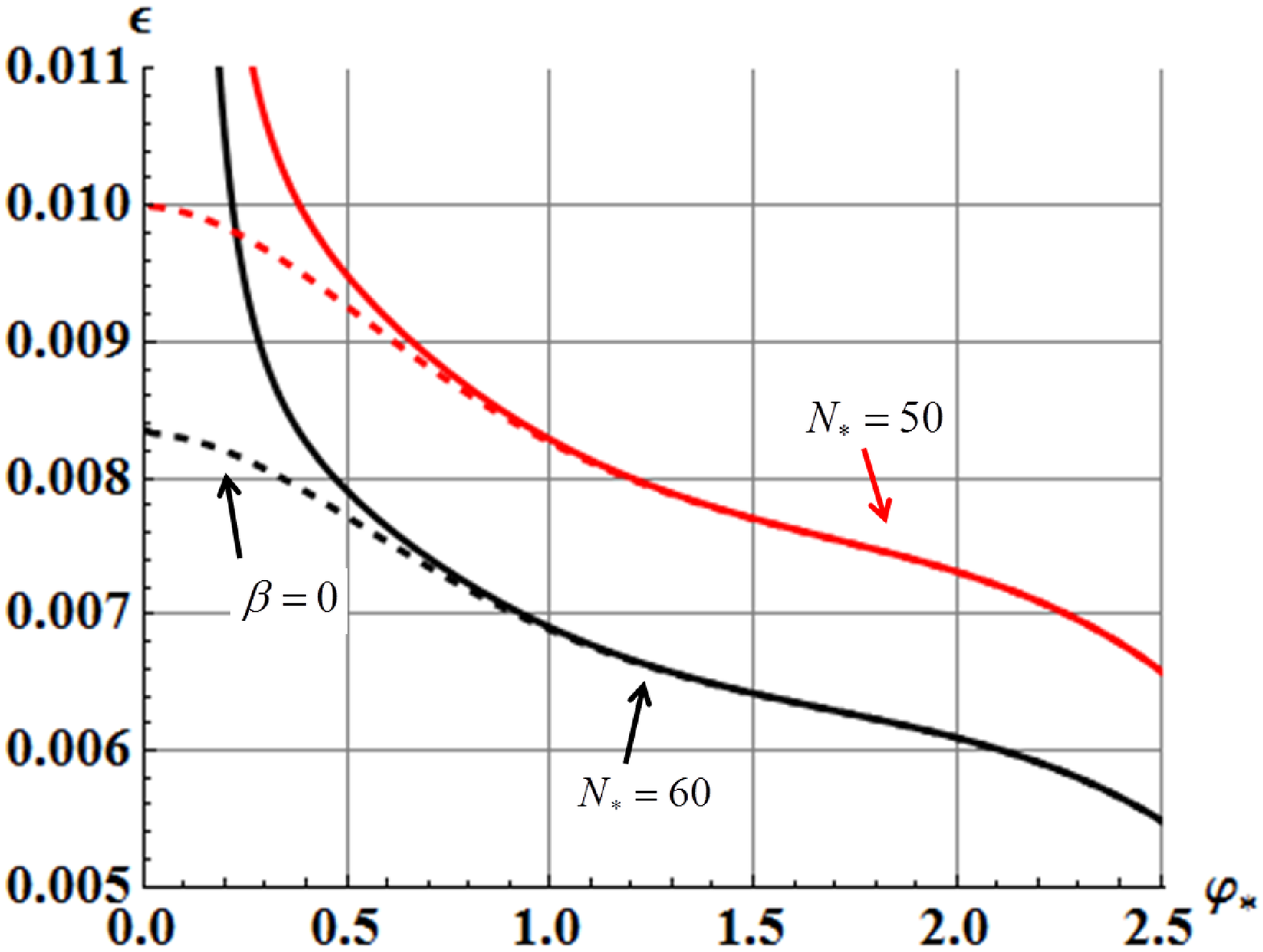,height=2.5in,width=3.3in}\hspace{0.5cm}
\epsfig{figure=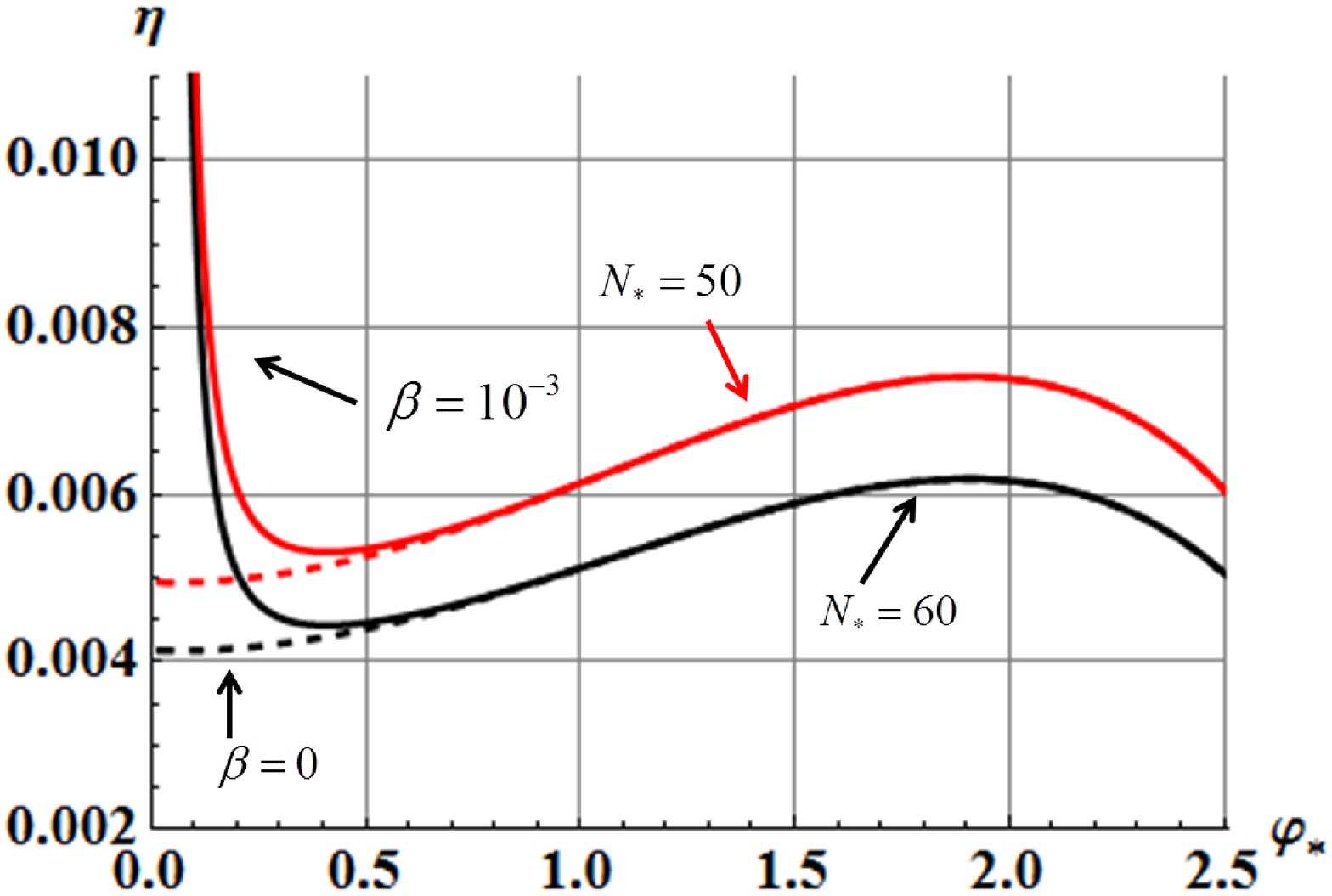,height=2.5in,width=3.3in}
\caption{The slow--roll parameter $\epsilon$ and $\eta$ with
$\beta=0$ (dashed line) and $\beta=10^{-3}$ (solid line).}
\end{center}
\label{fig-slow-roll}
\end{figure}

\medskip

\begin{figure}[ht]
\begin{center}
\epsfig{figure=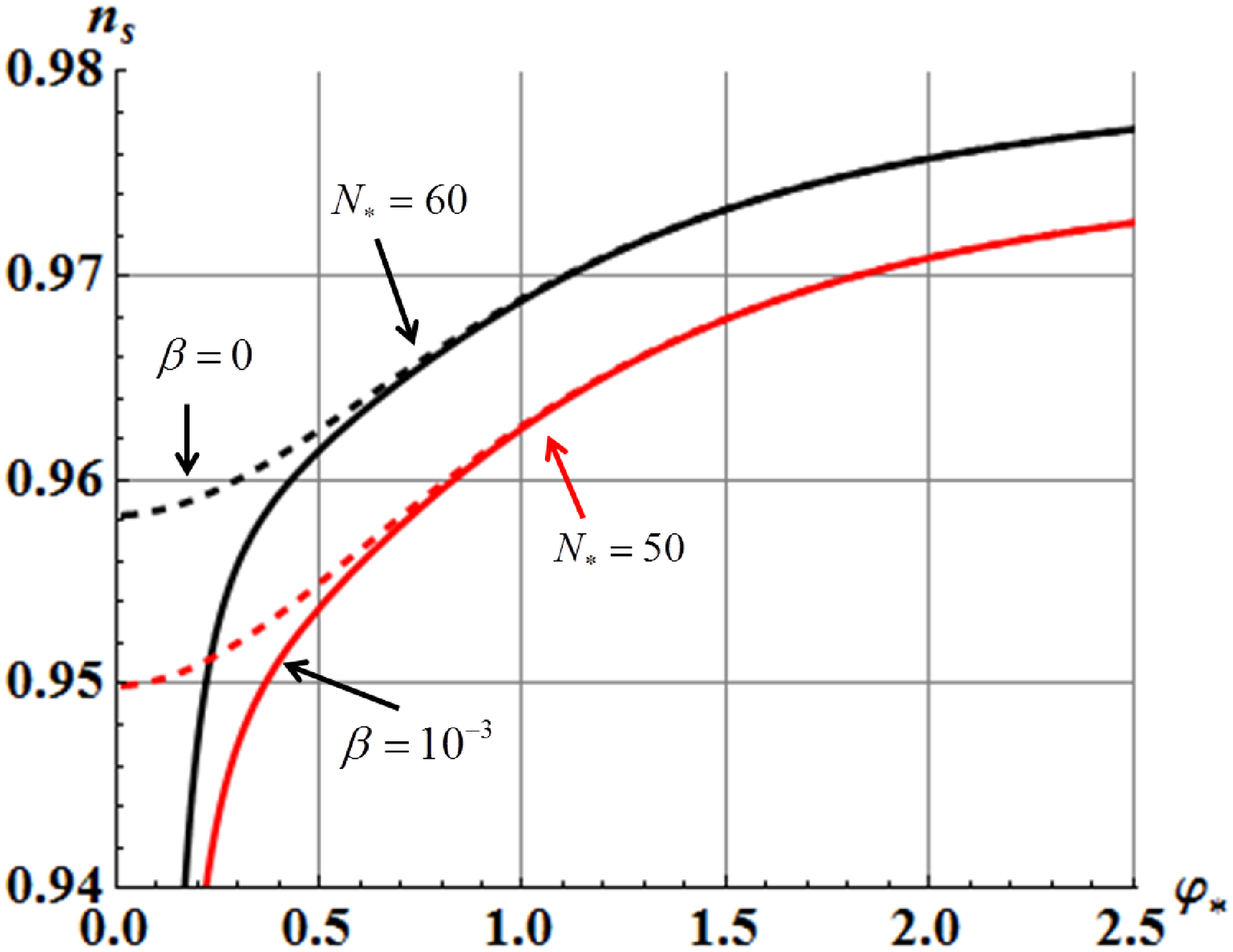,height=2.5in,width=3.3in}\hspace{0.4cm}
\epsfig{figure=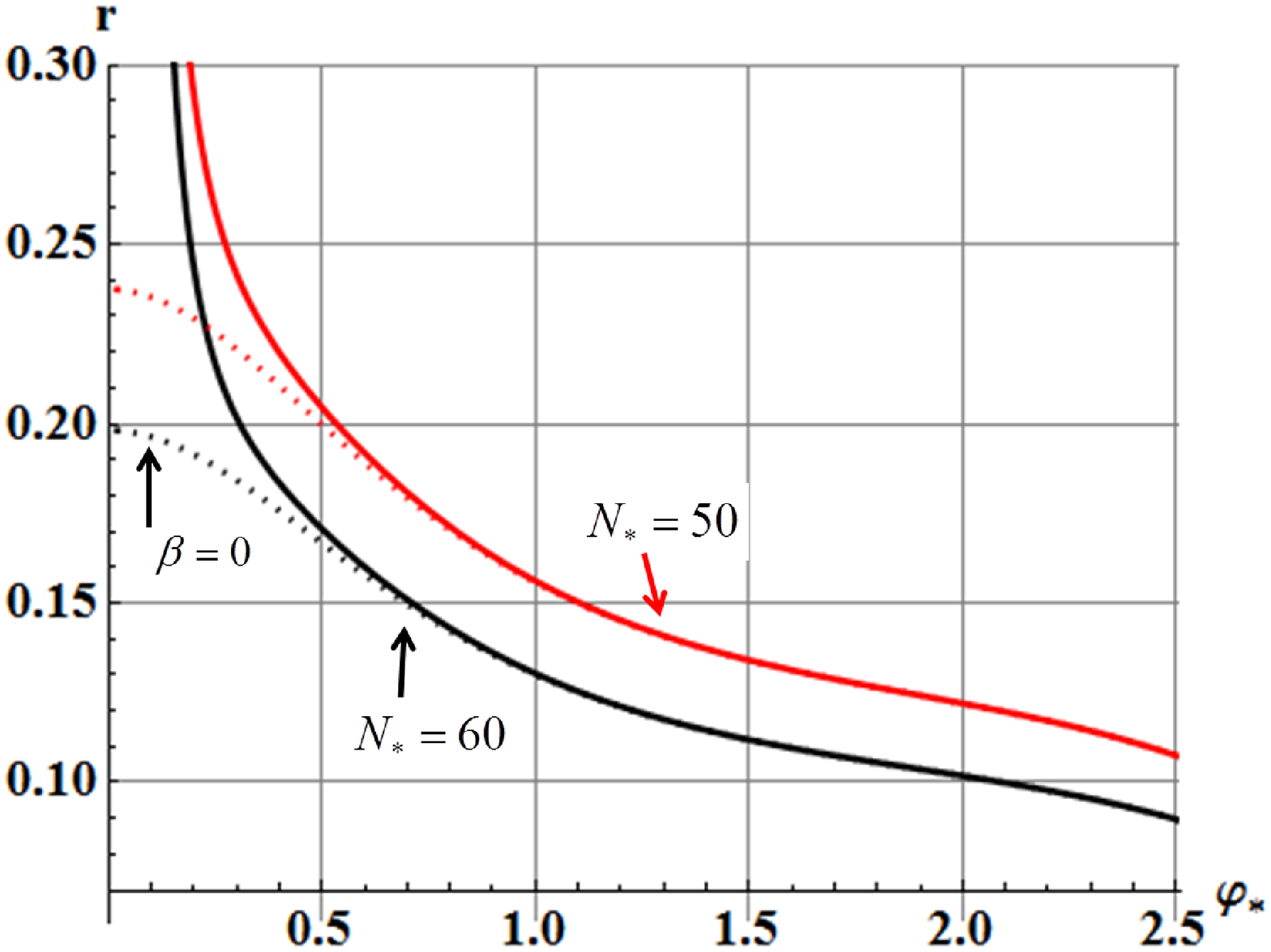,height=2.5in,width=3.3in}
\caption{The scalar spectral index $n_s$ and tensor-to-scalar
ratio $r$ with $\beta=0$ (dashed line) and $\beta=10^{-3}$ (solid
line).}
\end{center}
\label{fig-ns-and-r}
\end{figure}

\begin{figure}[ht]
\begin{center}
\epsfig{figure=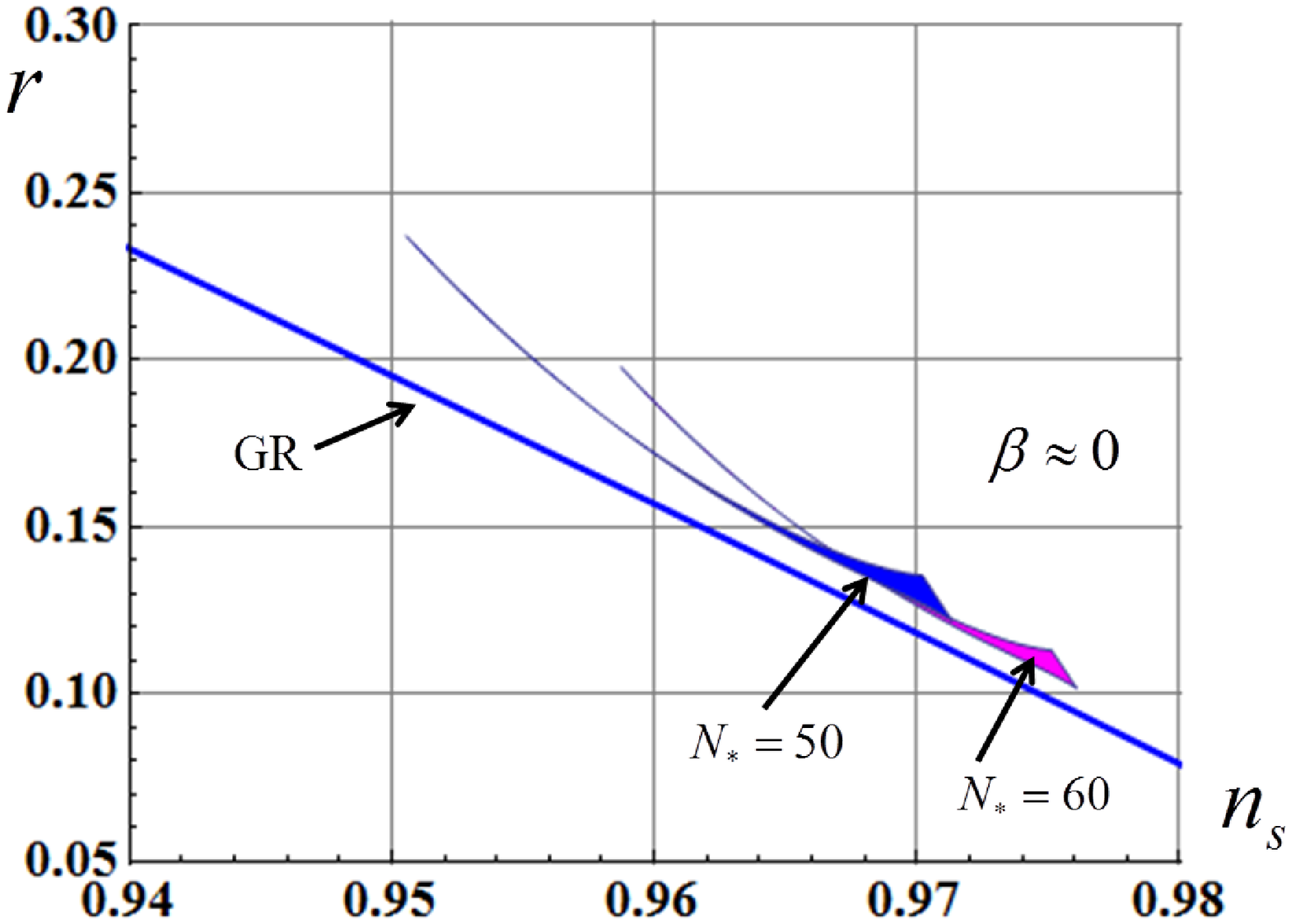,height=2.5in,width=3.2in}\hspace{0.4cm}
\epsfig{figure=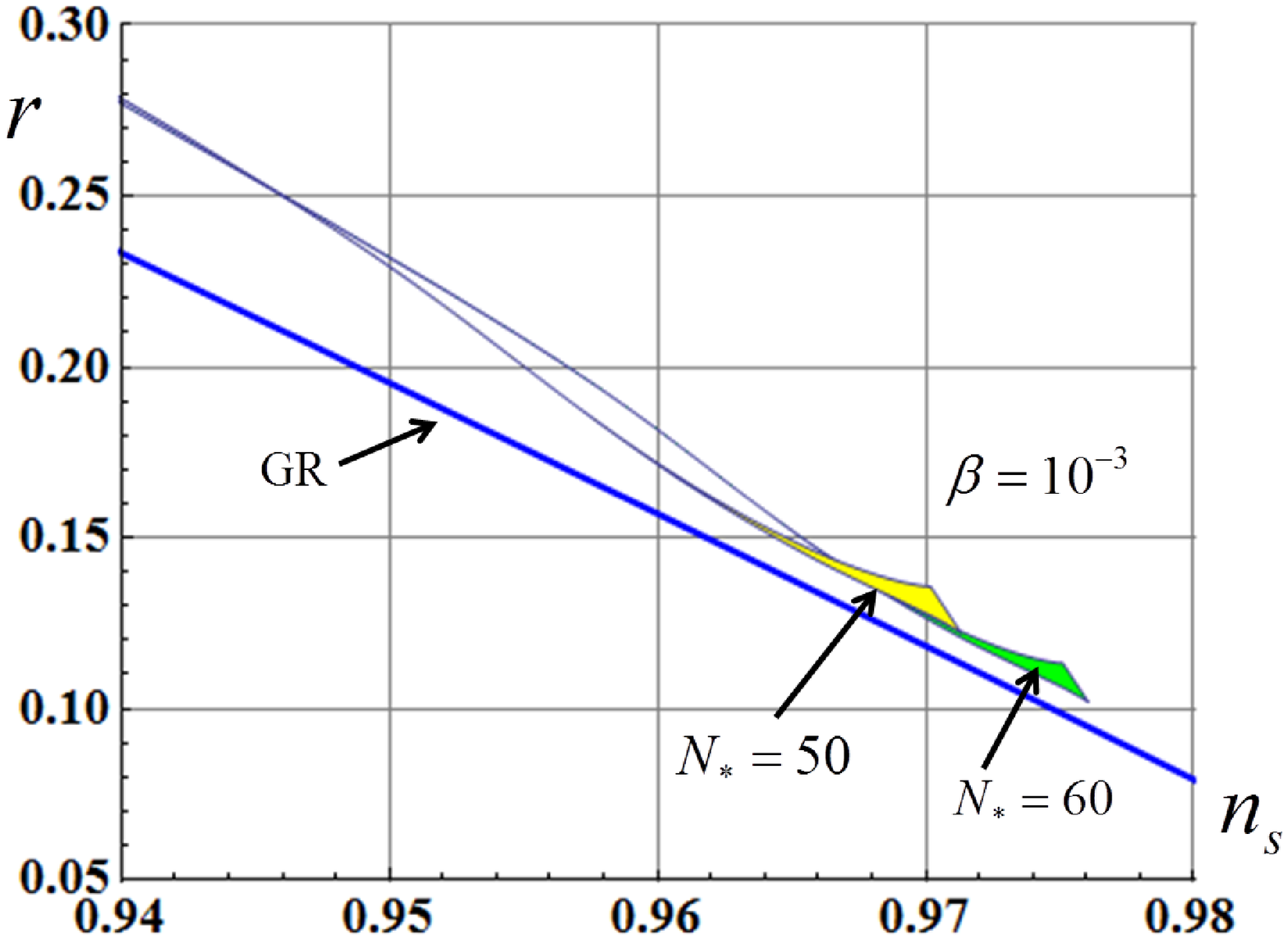,height=2.5in,width=3.2in} \caption{A
parametric plot: The tensor-to-scalar ratio $r$ versus the scalar
spectral index $n_s$ with $\beta=0$ (top plot) and $\beta=10^{-3}$
(bottom plot). $\varphi_*$ is varied from $\varphi_*=2$ to $0.05$.
For $\chi=0$, the cosine--form potential approximates to
$m^2\phi^2$ potential~\cite{Ish14a} and the shaded regions around
$n_s\sim 0.97$ are absent. The single solid line is the prediction
of $m^2\phi^2$ inflation~\cite{Creminelli2014} in 4D general
relativity.}
\end{center}
\label{fig-r-vs-ns}
\end{figure}

\medskip

As with a single-field, slow-roll inflation model in Einstein
gravity, on sufficiently large scales, we find that the growth of
scalar fluctuations depend on two parameters, $|\dot{\phi}|$ and
the Hubble scale $H$; more specifically, ${\cal P}_{\text
sca}^{1/2}\simeq H^2/(2\pi |\dot{\phi}|)$~\footnote{For cosmology
based on the Lagrangian~(\ref{main-5Daction}), almost all
contributions to primordial scalar fluctuations come from the
four-dimensional inflation field mainly because there are no
Kaluza-Klein excitations having a mass between $m^2=0$ and $m^2=
9H^2/4$~\cite{Maartens:2010,Neupane:2010}. The massive scalar
excitations with mass $m\Z{\rm KK}\ge 3H/2$ are rapidly
oscillating and their amplitudes are strongly suppressed on
sufficiently large scales~\cite{Maartens:1999,Neupane:2010}. This
feature is retained with ${\cal R}^2$ corrections to the
Lagrangian at least when $\sigma\ll V(\phi)$.}. The Hubble scale
$H$ is given by (\ref{main-Fried1}). During a slow-roll inflation,
since $\dot{\phi}\simeq - V_\phi/(3H)$, the amplitude of scalar
(density) perturbations is given by~\cite{Huey:2002,Dufaux:2004}
\be A\Z{S}^2\equiv {4\over 25} {\cal P}_{\rm sca}(k) \simeq
{9\over 25\pi^2} {H^6\over V_\phi^2}.\ee The normalized amplitude
of primordial tensor perturbations is given
by~\cite{Ish14a,Dufaux:2004,Sami04b,Bouhmadi-Lopez} \be A\Z{T}^2
\equiv {1\over 25} {\cal P}_{\rm ten}(k)= {2\over 25} {\psi\over
M^2 {\cal A}} \left({H\over 2\pi}\right)^2 ,\ee \be {\cal A}\equiv
(1+\beta)\sqrt{1+x^2} - (1-\beta) x^2 \sinh^{-1} {1\over x},\nn
\ee where $x\equiv H/(\psi M)=\beta^{-1/2} \left[(1-\beta)
\cosh\varphi-1\right]^{1/2}$ is a dimensionless measure of the
Hubble expansion rate. The power of the scalar and tensor
primordial spectra can be calculated approximately in the
framework of the slow-roll approximation by evaluating the above
equations at the value $\varphi=\varphi_*$. On the usual
assumption that $H$ is nearly constant throughout inflation, the
amplitude of scalar density perturbations has some scale
dependence due to a small variation in $V_\phi$, while the tensor
perturbations are roughly scale independent.

\medskip

The scalar spectral index is given by \be n_s-1 \equiv {d\ln
A_S^2\over d\ln k}\Big|_{k =aH}= -6\epsilon + 2\eta.\ee The
tensor-to-scalar ratio $r\equiv 4 {\cal P}_{\rm ten}/{\cal P}_{\rm
sca}$ is given by \bea r &=&{16\,I(\varphi_*) \over 2N_*+1}
{(1-\beta)^{3/2} |2\beta|^{1/2}\over {\cal A}} {\sinh(3\varphi_*/
2) \left(1-X(\varphi_*)\right)\over
\left[(1-\beta)\cosh\varphi_*-1\right]^2}.\nn \\\eea For example,
for $\beta\lesssim 0.003$, $N_*\sim (55-58)$ and $n_s\sim 0.96$
correspond to the values $\varphi_* \sim (0.570-0.425)$ and $r\sim
(0.177-0.181)$. In Figs. 2 and 3 we show the results in wider
ranges, $\varphi_*\sim 0.05-2.5$, $N_*=50-60$, and $n_s\sim
0.94-0.98$. The model leads to appropriate values for $H_*$ and
$V_*$ that are consistent with constraints from Planck data (see
below).

\begin{figure}[ht]
\begin{center}
\epsfig{figure=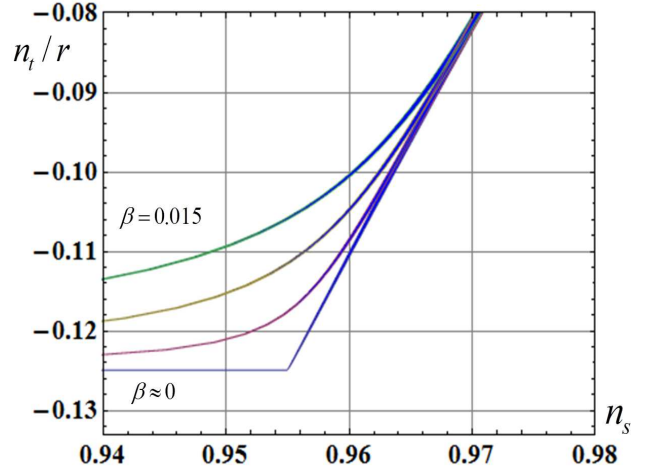,height=2.5in,width=3.3in}
\caption{The ratio $n_t/r$ versus scalar spectral index $n_s$ with
$N_*=55$ and $\beta=0.015, 0.005$, $0.001$ and $10^{-7}$ (top to
bottom). For $\beta\approx 0$, the ratio $n_t/r$ asymptotes to
$-0.125$ as $\varphi\to 0$ (general relativity limit).}
\end{center}
\label{nTbyr-ratio}
\end{figure}
\medskip

As an important consistency check of the model, we compute the
tensor spectral index: \be n_{t} = {d\ln A_T^2\over d \ln
k}\Big|_{k=aH}= -2\epsilon \times {1\over {\cal A}}\,{\beta x^2 +
\beta +1\over \sqrt{1+x^2} }.\ee 
In the limits $\beta\to 0$ and $x\to 0$, which means $H \psi \ll
M$, we recover the standard consistency relation that $
n_{t}=-2\epsilon$ and $n_{t}/r=-1/8$~\cite{Copeland:1993}, which
relate the tensor spectral index $n_t$ to the slow-roll parameter
$\epsilon$ and the ratio of the tensor and scalar perturbation
amplitudes. For the GB--assisted natural inflation model, we find
that the ratio $n_{t}/r$ always differs from the result in
Einstein gravity. In Fig. 4, we plot the ratio $n_{t}/r$ versus
the scalar spectral index $n_s$ for $N_*= 55$; by allowing the
coupling constant $\beta$ in a reasonable range ($0\ll
\beta<0.015$), we find that the ratio $n_t/r$ is in between $-
0.1002$ and $-0.1098 $ for $n_s\simeq 0.96$. This ratio, which
only modestly depends on $N_*$, is about (13\%--24\%) less than
the value predicted for models based on Einstein gravity. This is
one of the testable predictions of the model.

\medskip

Here we want to make a remark. Measuring $n_t$ may be challenging
with current technologies. However, if $r > 0.11$ as indicated by
the BICEP2 data ($r=0.16^{+0.06}_{-0.05}$ after subtracting an
estimated foreground), this might be feasible with the next
generation of space explorations~\cite{Andre:2013,Benson:2014}.
Recently, in~\cite{Easther2014b}, R. Easther {\it et al.} found
that the ratio $n_t/r$ is picked around $-0.15$ for a multifield
inflation characterized by the potentials $V\sim \sum_i \lambda_i
|\phi_i|^p$ with $p>3/4$, which differed from the prediction of
single-field slow-roll inflation by $5\sigma$ C.L. This prediction
is much larger than for single-field, slow-roll inflation in
Einstein gravity; a larger value of $n_t/r$ usually means a
smaller $r$, which seems contradictory to the value of $r$
reported by the BICEP2 experiment. The model proposed
in~\cite{Easther2014b} may be compatible with the BICEP2 data if
$|n_t|\gtrsim 0.024$. The Planck results put a constraint like
$|n_t|\le 2\epsilon \lesssim 0.02$.

\section{Observational constraints}

\begin{figure}[!ht]
\centerline{\includegraphics[height=2.4in,width=3.3in]
{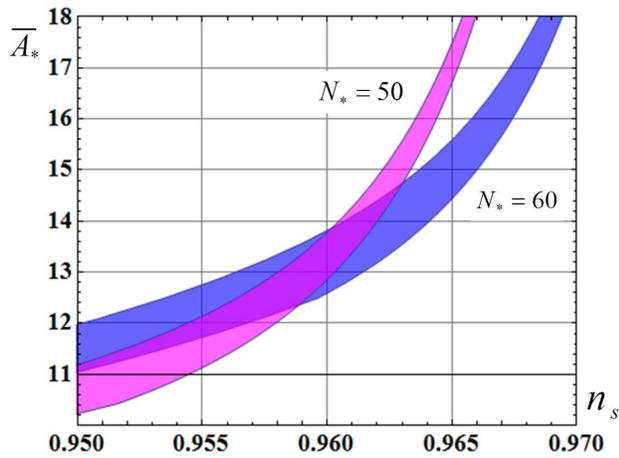}} \caption{The COBE normalized amplitude of scalar
perturbations $\bar{A}_*\equiv (M\Z{P}/M)^6 \times A_*$ versus
$n_s$ with $N_*=50$ and $N_*=60$ and $0< \beta
<0.01$.}\label{As-vs-ns}
\end{figure}
In order to constrain the model parameters, we use, as
in~\cite{Ish14a}, the COBE normalisation for amplitude of scalar
perturbations used by the Planck Collaboration~\cite{Planck2013},
$A_*\simeq V^3/(12\pi^2 M\Z{P}^6 V_\phi^2)\simeq 22\times
10^{-10}$, where $A_*$ is approximated as \be A_*= {2(2N_*+1)
\over 27\pi^2}{(1-\beta)^3\over I(\varphi_*)} {M^6\over M\Z{P}^6}
{\left[\sinh(3\varphi_*/2)\right]^2\over
\left(1-X(\varphi_*)\right)},\ee where, as usual, $\varphi_*$
denotes the value of $\varphi$ at the epoch at which a scale of
wave number $k$ crosses the Hubble radius during inflation. By
plotting $A_*$ versus the scalar spectral index $n_s$ (shown in
Fig. 5), we find that $n_s\simeq 0.9603$ and $N_*\sim 55$
correspond to the value
$$ A_* \simeq 13.6 \times \left(M/M\Z{P}\right)^6 \quad
\rightarrow \quad M\simeq 0.0233425\times M\Z{P}.$$ 
By using this result, along with the dimensional reduction
relation $ \psi M\Z{P}^2 = (1+\beta) M^2$~\cite{CNW-01} between
the four- and five-dimensional Planck masses, which holds as long
as $\beta$ and $\psi$ are constants~\footnote{ During inflation,
since ${\cal E}/a_0^4\approx 0$, $\beta$ and $\psi$ are
constants.}, one may express $\psi$ in terms of $\beta$ or vice
versa.

\medskip

The BICEP2 data appear to be consistent with the 2013 Planck
constrain on the scalar spectral index ($n_s \simeq 0.96$). The
COBE normalized number of {\it e}--folds (between the exit of
wavelengths now comparable to the observable universe and the end
of inflation) is $N\Z{\rm COBE}\sim 57$ (see below). By taking
these two observationally preferred values as input, we estimate
in Table I various quantities relevant to inflationary epoch or
inflation. This is a set of model parameters that lead to the
observationally preferred values of scalar spectral index $n_s\sim
0.96$ and the number of {\it e}--folds $N_*\sim 57$.

\begin{table}{Table I:} The set of parameters that lead to the
observationally preferred values of $n_s \sim 0.96$ and the number
of {\it e}--folds $N \sim 57$.
\begin{tabular}{ccccc}\hline\hline
$\beta$ & $\psi$ ($10^{-5}$) & $\varphi_*$ & $H_*$ ($10^{14}\,{\rm
GeV}$)
& $V_*^{1/4}$ ($10^{16}\,{\rm GeV}$) \\
\hline \hline $0.015$ & $ 55.26 $ & $0.775$ &$1.39$
 & 2.08 \\ \hline $0.010$ & $54.99$ & $0.720$ & $ 1.58 $ & $2.13$
\\ \hline $0.005$ & $54.72$ & $0.643$ & $2.00$ &
$2.24$ \\  \hline $0.003$ & $54.61$ & $0.595$ & $2.39$ & $2.33$
\\
\hline $0.001$ & $54.49$ & $0.529$ & $3.68$ & $2.58$
\\
\hline $0.00026$ & $54.46$ & $0.486$ & $6.64$ & $2.98$
\\
\hline\hline
  \end{tabular}
\end{table}\label{table1}

The numbers shown in Table I are tentative, which change if $N_*$
is found to be different from $N\Z{\rm COBE}$; if a deviation from
$N\Z{\rm COBE}$ is small, then the results are very similar.

\medskip

We can similarly constrain the model's parameters like $\Lambda$
and $f$. A small curvature coupling as $\beta\lesssim 0.015$ may
be sufficient for suppressing cubic and higher-order curvature
corrections in the Lagrangian and also radiative corrections; here
we allow $\beta$ in a slightly wider range $10^{-6}< \beta< 0.02$.
With $N_*\simeq 55$ and $n_s\simeq 0.9603$, and using the
condition~(\ref{matching}), we observe that \be 21.56 \le \xi^2
\le 23.12, \quad \xi^2 \equiv {10^4}\times {4\alpha\times
\Lambda^4 \over M^2 f^2}. \label{bound-nat-inf}\ee The smaller the
GB coupling is, the larger the ratio $\Lambda/\sqrt{M f}$ would
be. If we take the value $f\sim M\Z{P}$ and $\Lambda\sim 1.0\times
10^{16}~{\rm GeV}\equiv \Lambda_*$ as motivated in string--theory
models~\cite{Freese:1990,Adams:1992} or by CMB
observations~\cite{BICEP2,Planck2013}, then we find that $ \alpha
\sim 1024- 1098$ or vice versa.

\medskip

Of course, the bound (\ref{bound-nat-inf}) alters once the number
of {\it e}--folds is changed; specifically, with $N_*=50-60$ and
$n_s=0.96$, we have $4 < \xi^2 < 60$. Similarly, a deviation from
$n_s\simeq 0.96$ also changes the bound. For $n_s = 0.9603 \pm
0.0073$ (which is within 68\% or 1$\sigma$ confidence level result
of Planck 2013 data) and
$N_*=50-60$, the bound on $\xi^2$ is given by 
\be 3 < \xi^2 < 150. \label{wide-bound}\ee If $\beta$ is closer to
zero then $\xi^2$ is closer to the lower limit. If $\alpha\simeq
0$ then one would require a much larger value for $\Lambda$ that
is inconsistent with an upper bound on the energy scale of
inflation, $V_*^{1/4}=1.94\times 10^{16}~{\rm
GeV}\,(r_*/0.12)^{1/4}$~\cite{Planck2013}. The $\alpha=0$ case is
ruled out; inflation based on Randall-Sundrum
cosmology~\cite{RS2,Okada:2014} cannot explain the observational
bound on the energy scale of inflation. One would require $\beta>
0.0001$ (and hence $\alpha\gtrsim 180$) for consistency of the
model with Planck results. Indeed, in the present context, the
value of the Gauss-Bonnet coupling is very important for a
determination of the energy scale of cosmic inflation or vice
versa.

\medskip

In Fig.~6, we plot the function $\xi^2$ by varying the GB coupling
in the range $\alpha\sim (10^{2}-10^{4})$. For a larger $\alpha$,
the ratio $\Theta$ is smaller. As the plot shows, the value of
$\Theta$ may be allowed anywhere between $1$ and $100$, which
translates to the bound \be 0.006132 < {\Lambda/\sqrt{M\Z{P} f}} <
0.01.\label{bound-Lf}\ee For $f\sim 0.68 \times M\Z{P}$, as
motivated in string-theory models, we find that $\Lambda\sim (0.63
- 2.0)\times 10^{16}\,{\rm GeV}$. This bound is fully consistent
with an upper limit on the energy scale of inflation from Planck
data $V_*^{1/4} \lesssim 2.37\times 10^{16}\,{\rm GeV}$ (for
$r_*<0.27$) or $\Lambda< 2.17\times 10^{16}\,{\rm GeV}$. The
smaller is the value of $f$ (with $f< M\Z{P}$), the narrower would
be the bound for $\Lambda$, which is desirable both theoretically
and observationally. A similar bound on $V_*^{1/4}$, namely
$V_*^{1/4} \sim (2.07-2.40)\times 10^{16}~{\rm GeV}$, was obtained
in~\cite{Choudhury:2014} imposing that $r_*\sim 0.15-0.27$.

\begin{figure}[!ht]
\begin{center}
\epsfig{figure=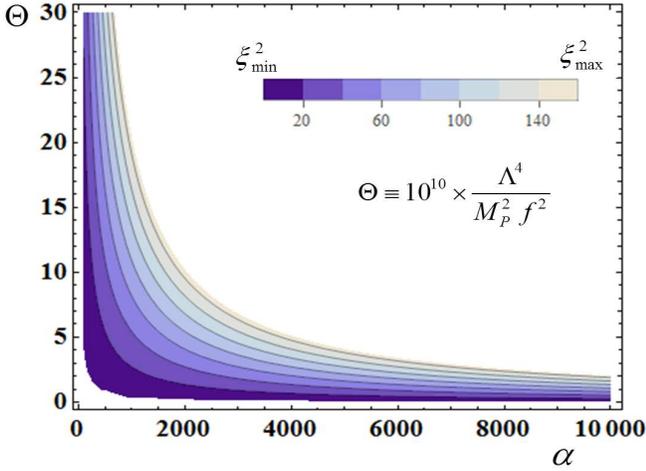,height=2.5in,width=3.4in}
\caption{The bound $3<\xi^2<150$ as a function of $\Theta$ and the
GB coupling $\alpha$.}
\end{center}
\end{figure}

\medskip

Note that $R\propto H^2$ and ${\cal R}^2\propto H^4$. This
implies, for example, if $\alpha\sim 10^4$, the Gauss-Bonnet term
$\alpha ({\cal R}^2/M^2)$ is subleading to the Einstein-Hilbert
term for $H/M< 10^{-2}$. In fact, the Planck data put an upper
bound on the Hubble scale of inflation, namely, $H_*< 8.8\times
10^{14}~{GeV}$. So, with $M\sim 5.67\times 10^{16}\,{\rm GeV}$,
the ${\cal R}^2$ term is subleading to the Einstein-Hilbert term
for $\alpha\lesssim 5\times 10^{3}$. A larger $\alpha$ than this
may be allowed if $H < H_*$.

\medskip

The recent analysis of Planck+WAMP-9+high L+BICEP2 data leads to
slightly modified bounds, namely $V_*^{1/4}=2.4\times 10^{16}~{\rm
GeV}\,(0.27/r(k_*))^{1/4}$ and $0.15 < r(k_*)< 0.27$ at the pivot
scale, $k_* = 0.002 {\rm M pc^{-1}}$.

\medskip

The above estimate is only tentative since the results depend on
the ultimate values of $N_*$ and $n_s$. Nevertheless, the numbers
are quite impressive in the sense the GB--assisted ``natural
inflation" is in perfect agreement with {\textit Planck} data for
a reasonable range of the energy scale of inflation, number of
{\it e}--folds and scalar spectral index. The observation that
GB-assisted natural inflation parametrized by the potential
(\ref{PNGB}) is consistent with the Planck bound on the energy
scale of inflation $V_*^{1/4}$ and also with the recent BICEP
result $r_*=0.19^{+0.007}_{-0.005}$ with $f \lesssim M\Z{P}$ is
quite remarkable. For values of $f$ sufficiently near $M_{\rm
Pl}$, sufficient inflation takes place for a broad range of
initial values of the field $\phi$.

\subsection*{Limits on $ V_*^{1/4}$ and shift in $\phi$}

\begin{figure}[!ht]
\centerline{\includegraphics[height=2.5in,width=3.4in]
{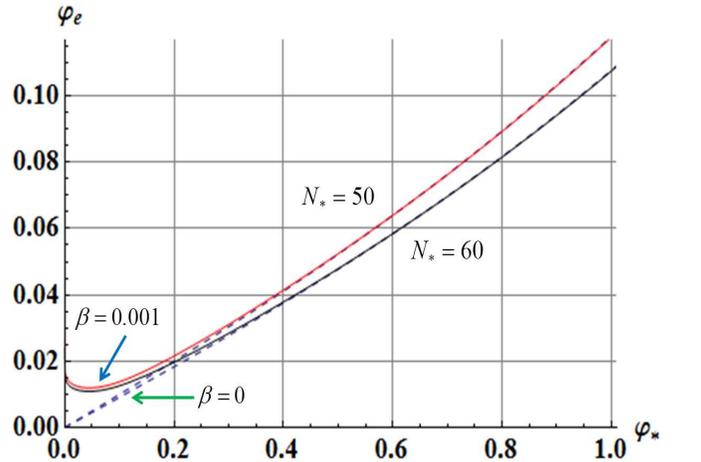}} \caption{The $\varphi_e$ as a function of
$\varphi_*$ for $\beta=0$ (dashed lines) and $\beta=0.001$ (solid
lines).}\label{fig-phiend}
\end{figure}

For $N_*\sim 55$, the scalar spectral index $n_s \sim
0.9603^{+0.005}_{-0.005}$ corresponds to the tensor-to-scalar
ratio to $r_*= 0.176^{-0.028}_{+0.039}$ and to the field value
$\varphi_*\sim 0.31-1.10$. As shown in Figure~\ref{fig-phiend} the
variation $\varphi_*\sim (0.31-1.1)$ implies $\varphi_e\sim (0.03-
0.12)$. It follows that $\Delta\varphi\equiv
\varphi_*-\varphi_{\rm end}$, the change in $\varphi$ after the
scale $k_*$ leaves the horizon, $\Delta\varphi \sim (0.28 -
0.98)$, depending upon the energy scale of the inflation. This
implies \be V_{\rm end}^{1/4} \sim 0.55 \times
V_*^{1/4}.\label{Vin-vs-Vfin}\ee The slow-roll condition is well
satisfied, which guarantees the existence of an inflationary
epoch. The condition $V_*\gg \sigma$ is also justified. Typically,
if
$$ {\phi_*}\sim \pi f, \quad {\rm then} \quad {\phi_{\rm
end}}\sim 5f/3.$$ Likewise, if $\phi_*/f\sim 2$ then $\phi_{\rm
end}/f\sim 1.35$, which means $\Delta\phi= \phi_*-\phi_{\rm end} <
1.47\,f$. The distance traversed by the inflaton field $\phi$ is
always sub-Planckian as long as $f < 0.68\,M\Z{P}$, which means
the trans-Planckian
problem~\cite{Martin:2000,Kim:2004,Mazumdar:2014} is absent. This
is a direct consequence of the fact that the ${\cal R}^2$
corrections ease the slow-roll conditions for inflation and enable
inflation to take place at field values below $M\Z{P}$. This is a
very important result in view of the earlier observation (in
conventional GR) that the model agrees with Planck+WP data for $f
> 5 M\Z{P}$~\cite{Planck2013}.
The above conclusion is qualitatively the same for $N_*\sim
50-60$.

\medskip

The above results also apply to GB-assisted $m^2\phi^2$ inflation
(with $m \equiv \Lambda^2/(\sqrt{2} f)$ as it is a limiting case
of GB-assisted natural inflation, especially, around and below the
energy scale of inflation, $\varphi_* \lesssim 1.1$. The
prediction for running scalar spectral index in natural inflation
may be different from that in the case of chaotic inflation. Near
future observations from Planck experiments for the running
spectral index may achieve enough accuracy to allow us to
distinguish GB--assisted natural inflation from GB--assisted
chaotic inflation.

\section{Reheating of the Universe}

Once $\phi$ rolls (roughly) below $0.1\, f$, or when $\varphi\ll
\varphi_{\text end}$, the field evolution may be described in
terms of oscillations about the potential minimum. For small
enough amplitude, the potential is well approximated by $V(\phi)=
m^2 \phi^2$ with $m^2\equiv (\Lambda^4/2 f^2)\sim (9.6\times
10^{13}\,{\rm GeV})^2$ for $\Lambda\sim 1.5\times 10^{16}\,{\rm
GeV}$ and $f\sim 1.65\times 10^{18}\,{\rm GeV}$ (to be roughly
consistent with the normalization of the
power spectrum discussed in the above section). 

As in {\it natural inflation} and $m^2\phi^2$-inflation scenarios
in Einstein gravity, at the end of the slow-rolling regime, the
field $\phi$ oscillates about the minimum of the potential and
gives rise to particle and entropy production. The cold
inflaton-dominated universe can undergo a phase of reheating once
the field value drops well below $0.1 M\Z{P}$, during which the
inflaton decays into ordinary particles and the Universe becomes
radiation dominated. The reheating temperature may be approximated
by  \bea T\Z{\text RH} & \sim & V\Z{\text end}^{1/4}
\left({\Gamma\over M}\right)^{1/2}\sim \left({45\over 4\pi^2
g_*}\right)^{1/4} \left(\Gamma
M\Z{P}\right)^{1/2},\label{Reheat-NI}\eea where $\Gamma$ is the
decay rate of the $\phi$ field into light fermions (or gauge
bosons) and $g_*$ is the number of relativistic degrees of
freedom. In the above result we used the approximation $V_{\text
end}\sim 4\times 10^{-2} M^4/\sqrt{8\alpha}$, $M\sim 0.0233
\,M\Z{P}$, and $\sqrt{\alpha}\sim 0.08\times g_*$, so that it
closely resembles with the result obtained by Adams {\it et.al.}
in~\cite{Adams:1992}. Here we are only trying to make a rough
estimate of $T\Z{\text RH}$, so the precise value of $g_*$ or the
GB coupling does not make a big difference to any of the
statements below. On dimensional grounds, the decay rate is given
by (see, for example,~\cite{Shtanov:1994})
$$ \Gamma\simeq \Upsilon^2 \times m_\phi,$$ where
$\Upsilon \equiv g \times (m_\phi/ f)$ is the Yukawa coupling and
$g$ is an effective coupling constant. This approximation is valid
not just in a Minkowski space but also in an expanding
universe~\cite{Shtanov:2014}, provided that $H\ll m_\phi$ (during
reheating). Hence \be \Gamma \simeq g^2 {m_\phi^3\over f^2} = g^2
{\Lambda^6\over f^5},\ee where we used $m_\phi^2 \equiv
V_{\phi\phi}\vert_{\phi=0}=\Lambda^4/f^2$.
Equation~(\ref{Reheat-NI}) reads as  \be T\Z{\rm RH} \sim
{0.35\over \alpha^{1/8}} \times \left({M\over f}\right)^{1/2}
{g\Lambda^3\over f^2}.\ee For example, if we take $f\sim 0.6
M\Z{P}$ and $\Lambda= 1.51\times 10^{16}~{\rm GeV}$ [which is well
inside the bound defined by Eq.~(\ref{bound-Lf})] and $g \sim
0.1$, then tentatively we find that $T\Z{\rm RH}\simeq 1.1\times
10^{10} \times \alpha^{-1/8}~{\rm GeV}$. For example, for
$\alpha\sim 10^2$, this yields $T\Z{\rm RH}\sim 6.2\times
10^{9}~{\rm GeV}$, which is physically viable.

\medskip

The number of {\it e}--folds between the exit of wavelengths now
comparable to the observable universe and the end of inflation, or
the COBE normalized number of {\it e}--folds is \be N_{\rm COBE}
\sim 62 -\ln {10^{16}~{\rm GeV}\over V_*^{1/4}}+ \ln
{V_*^{1/4}\over V_{\rm end}^{1/4}}- {1\over 3} \ln {V_{\rm
end}^{1/4}\over \rho\Z{\rm RH}^{1/4}},\ee where $\rho\Z{\rm RH}$
is the energy density in radiation as a result of reheating. If
$\rho\Z{\rm RH}^{1/4} \sim T_{\rm RH}$, then using
(\ref{Vin-vs-Vfin}) we find $N_{\rm COBE} \sim 57$. The spectral
index approximated by $n_s \sim 1-2/N\Z{\rm COBE}$ is $n_s\sim
0.965$ -- a value which is well within $1\sigma$ confidence level
(68\%) of the Planck data. This shows the consistency of the
model, independent of a bound on $r$.

\medskip

A theory of baryogenesis during reheating discussed, for example,
in~\cite{Dolgov:1996,Kofman:1997} (see
also~\cite{Felipe:2004,Copeland:2005}) to explain how particle
production after the end of inflation can be applied to the
present model. As in the standard natural inflation model,
baryogenesis can take place mostly during the reheating era,
whereas nucleosynthesis can take place at a later stage but well
before the Universe enters into a late--epoch cosmic acceleration
-- the second epoch of cosmic inflation but at a much slower pace.
The inflaton density can drop significantly after a period of
parametric resonance (or during the phase of coherent
oscillations). The Universe can decelerate all the way until
$\rho^{1/4}$ drops below $238 \,{\rm GeV}$. A detailed theory of
baryogenesis would require a deeper understanding of particle
physics around the energy scale of reheating, such as, the effects
of various interactions between $\phi$ and fermions and the other
decay products of $\phi$ (or bosons). This topic is beyond the
scope of this paper.

\section{Dark energy cosmology}

In this section we establish that the model may be used to explain
the concurrent universe with the right amount of dark energy
equation of state and the present Hubble scale.

\subsection{Low energy limit}

At low energies, $\varphi=\varphi_0 \ll 1$. Expanding around
$\varphi_0 =0$, we find that \bea H^2 = {M^2\psi_0^2\over \beta_0}
\left[(1-\beta_0)\left(1+{\varphi_0^2\over 2} + \cdots
\right)-1\right],\eea 
\bea \varphi_0 \approx {2\over 3} {(\rho+\sigma)\over \psi_0
M^4}{(2\beta_0)^{1/2}\over 4(1-\beta_0)^{3/2}}, \eea where
\be \beta_0 \equiv 4\alpha \psi_0^2= 1 - \left(1+8\lambda \alpha +
{8\alpha {\cal E}\over a_0^4 M^2}\right)^{1/2}.\ee In principle,
$\rho= \rho\Z{\text M}+ \rho\Z{\text R}+\rho\Z{\phi}$ but the
inflaton contribution can be negligibly small at late epochs since
$\phi=0$ is a minimum for a cosine--form potential, which means
$\rho\simeq \rho\Z{\text M}+ \rho\Z{\text R}$. After inflation
(more precisely, after reheating), the bulk radiation term
proportional to ${\cal E}(a_0)$ is nonzero. The GB coupling
$\alpha=\beta_0/(4\psi_0^2)$ is assumed to be a
constant.~\footnote{As a variant of this idea, one may allow a
slowly varying $\alpha$ as in string theory where $\alpha$ is
proportional to the Regge slope $e^{\bar\phi}/g_s^2$, where
$\bar\phi$ is the dilaton and $g_s$ is the string coupling.}

\medskip

For $\rho+\sigma \ll M^4$ and $\beta_0 \ll 1$, the Friedmann
equation reduces to \bea H^2 = H_0^2 + {\rho^2 \over 36
(1-\beta\Z{0})^2 M^6} + {\sigma \rho \over 18 (1-\beta\Z{0})^2
M^6},\label{main-Fried-GB}\eea where \bea H_0^2 &\equiv & - M^2
\psi_0^2 + {\sigma^2\over 36 (1-\beta_0)^2 M^6}\nn \\
& \simeq & - M^2 \psi_0^2 + {\sigma^2\over 36 M^6}+
{\beta_0\sigma^2\over 18 M^6}.\label{def-H02}\eea On large scales,
$c H_0^{-1} \sim 1.3\times 10^{28}\,{\text cm}\sim
4222\,\text{Mpc}$, the proportion of dark energy and (ordinary
plus dark) matter appear to be 68.3\% and 31.7\% at present, which
means \bea \Omega_{m} &=& {\sigma \rho\over 18(1-\beta_0) M^6
H^2}\left(1+
{\rho\over 2\sigma}\right)\sim 0.317, \nn \\
\Omega_\Lambda &\equiv& {H_0^2\over H^2} \sim 0.683.\eea Moreover,
$\rho\gtrsim \rho_c = 3.98 \times 10^{-47}~\text{GeV}^4$, which
means $\sigma^{1/4} \lesssim 1.09\times 10^{16}\,\text{GeV}$. This
is not surprising because the brane tension is large when the size
of the Universe is also large, which is actually proportional to
the volume of the Universe,~\footnote{If we take a smaller patch
of the Universe, then the brane tension is also small. For
example, on galactic distances, the value of $\sigma$ can be much
smaller than its value on Hubble scales.}. There also exists a
lower bound on the 3-brane tension (see below). Here we must note
that $\sigma$ is a free parameter and the RS-type fine-tuning of
brane tension, $\sigma=2 M^4\psi_0 ( 3-\beta_0)\sim 6 M^4 \psi_0$
(since $\beta_0\simeq 0$) holds only when $\rho=0$, $H=0$ and also
${\cal E}=0$, but {\it not} if any of these quantities is not
zero.

\medskip

As in the $\alpha=0$ case~\cite{Ish012}, the Universe can undergo
transition from decelerating to accelerating expansion when \be
 {\text w}_{\rm{eff}} = {p-\sigma\over \rho+\sigma} = {{\text
 w}-\xi\over 1+\xi} \lesssim
-1/3,\ee where $\xi\equiv \sigma/\rho$ and ${\text w} = p/\rho$ is
the equation of state of matter or radiation. For example, for
$\xi=200$, we obtain ${\text w}_{\text {eff}} \simeq - 0.995$,
which is indistinguishable from the effect of a pure cosmological
constant. Acceleration kicks in first on the largest scales as the
condition $\rho\ll \sigma$ is achieved there at first. It should
be noted that the condition $\rho< 2\sigma $ is {\it not} always
sufficient for the occurrence of cosmic acceleration; it also
depends on the relative ratio $\rho\Z{\Lambda}/\rho$ or the ratio
$\nu\equiv \sigma/(6H_0 M^3)$ (in the present model). For example,
a domain of spacetime with $\nu\gg 1$ does not enter into an
accelerating phase unless that $\xi\gg 1$ is attained.

\subsection{Late epoch acceleration}

In the post-inflationary universe it is natural to assume that the
energy density decays as
\begin{equation}
\rho={\rho_{*}\over a^{\gamma}},\qquad\gamma=3\left(1+\text{w}
\right),\label{cons-rule} \end{equation}
where $\rho_*$ is a constant and $\gamma=3$ ($\gamma= 4$) for
ordinary matter (radiation). Equation~(\ref{main-Fried-GB}) admits
an exact solution, which is given by
\begin{eqnarray}
 &&a^\gamma = {{\rho}_{*}\,\nu \over \sigma (1-\beta_0)^2 }\nn\\
&&  \times \Big[(1-\beta_0)\sinh(\gamma H\Z{0} t) + \nu
\left(\cosh( \gamma H\Z{0}t)-1\right)\Big], \label{sol-scale2}
\end{eqnarray} where $\nu\equiv {\sigma/ (6 H\Z{0} M^3)}$. The
matter (radiation) density evolves as
\begin{eqnarray}
\rho = \left({\sigma \over \nu}\right) {(1-\beta_0)^2\over
{(1-\beta_0)\sinh(\gamma H\Z{0} t) + \nu \left(\cosh( \gamma
H\Z{0}t)-1\right)}}. \label{Density}\nn \\ \end{eqnarray} In the
limit $\beta_0\to 0$, we recover the results in
\cite{Ish012,BDL2}. The Hubble expansion parameter and
deceleration parameters are obtained by using the definition
$H:=\dot{a}/a$ and $q = - 1 - \dot{H}/H^2$. In fact, $H_0=0$ is
{\it not} a physical choice, so we take $H_0>0$. It is readily
seen that the scale factor grows in the beginning as
$t^{1/\gamma}$ but at a late epoch it grows almost exponentially,
\bea a(t) &\simeq & \left({\rho_* \nu \over 2\sigma(1-\beta_0)^2}
\right)^{1/\gamma} \Big( (1-\beta_0+\nu)\,e^{\gamma H_0 t}-
{2\nu}\Big)^{1/\gamma}\nn \\
& =& \left({\rho_*\over \sigma}\right)^{1/\gamma} \Big( e^{\gamma
H_0 t}- 1\Big)^{1/\gamma},\eea where the equality holds in the
limit $\nu \to 1$ and $\beta_0\to 0$. The result shows that after
the end of inflation (more precisely, after reheating) the scale
factor could grow much slower than that predicted by Einstein
gravity; specifically, $a\propto t^{1/4}$ ($\propto t^{1/3}$)
during radiation (matter) dominated era. The period of structure
formation can be longer than in GR.

\begin{figure}[ht]
\begin{center}
\epsfig{figure=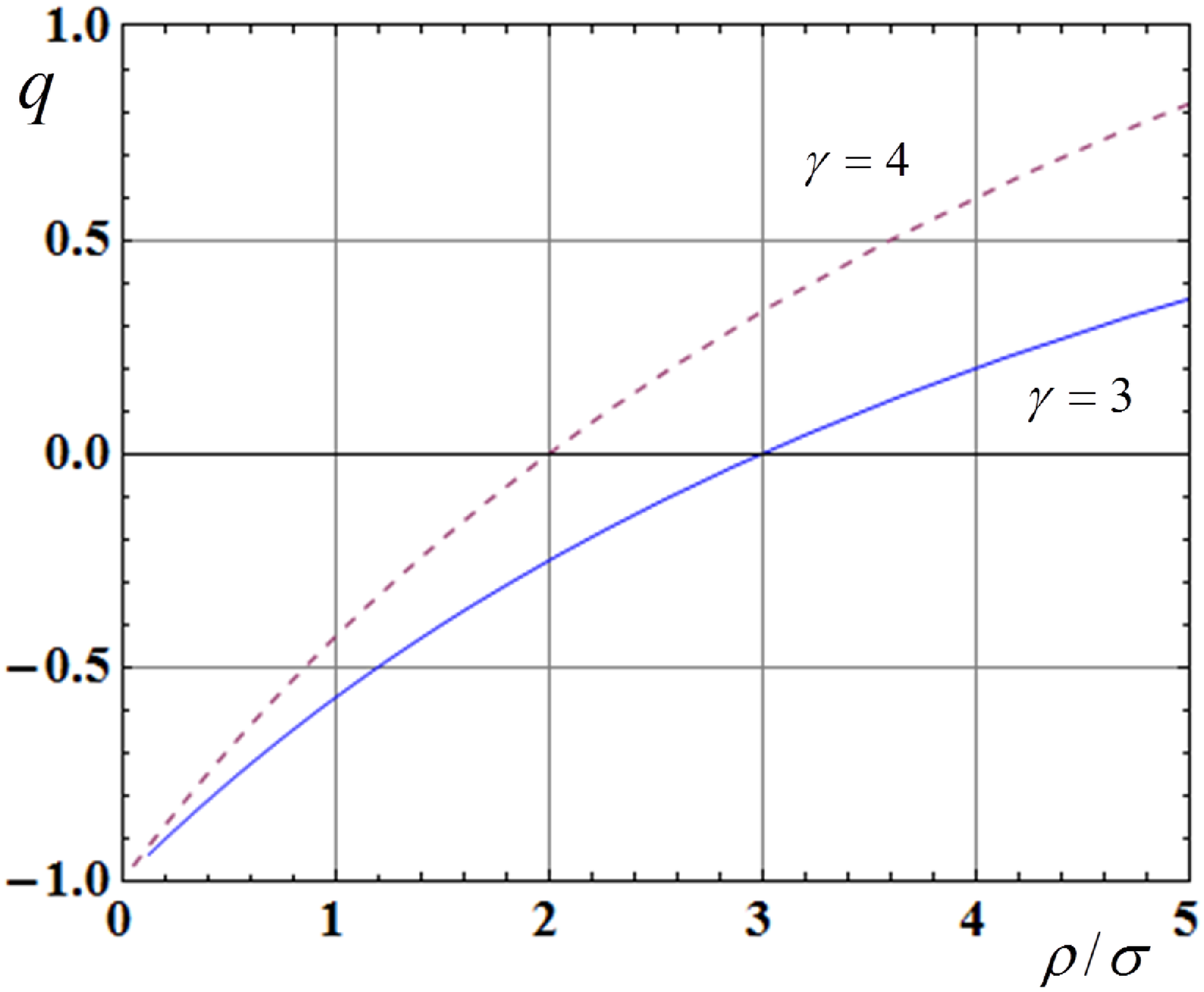,height=2.4in,width=3.1in}
\epsfig{figure=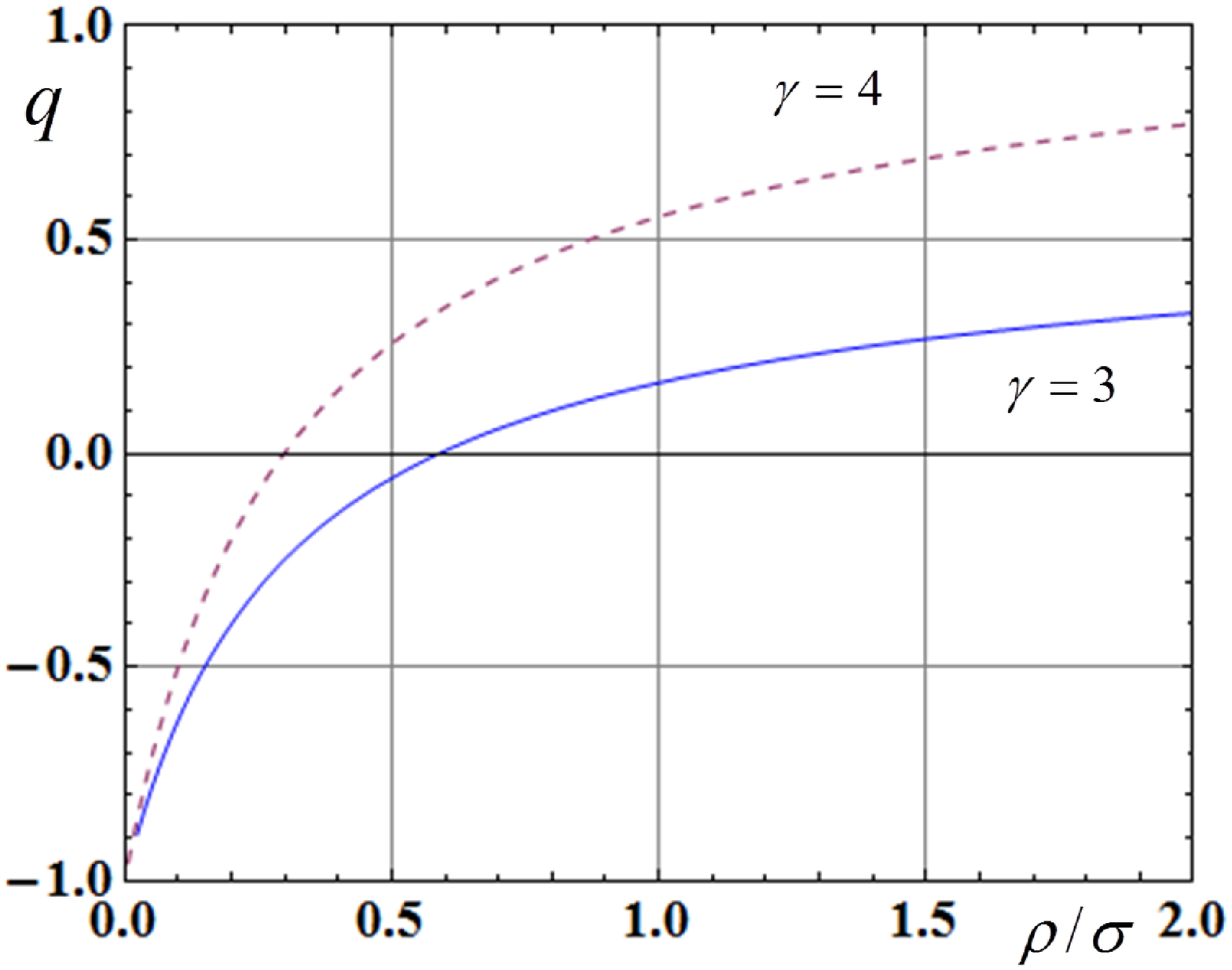,height=2.4in,width=3.2in}
\caption{A parametric plot: Deceleration parameter $q$ versus the
ratio $\rho/\sigma$ with $\nu =1$ (top plot) and $\nu=10$ (bottom
plot). $H_0 t$ is varied from $0.01$ to $1.5$ and $\beta_0 \gtrsim
0$.}
\end{center}
\end{figure}
\begin{figure}[ht]
\begin{center}
\epsfig{figure=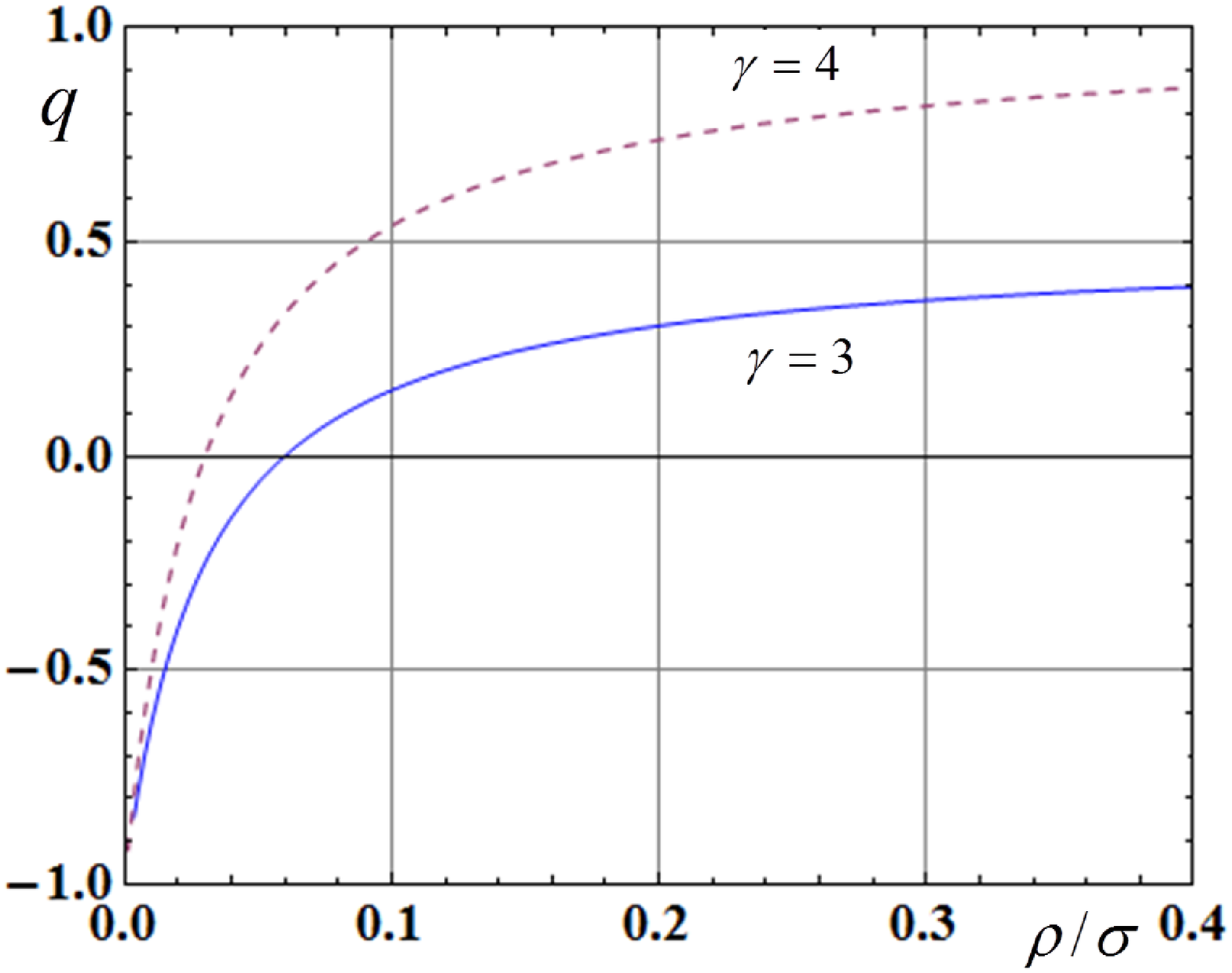,height=2.4in,width=3.2in}
\epsfig{figure=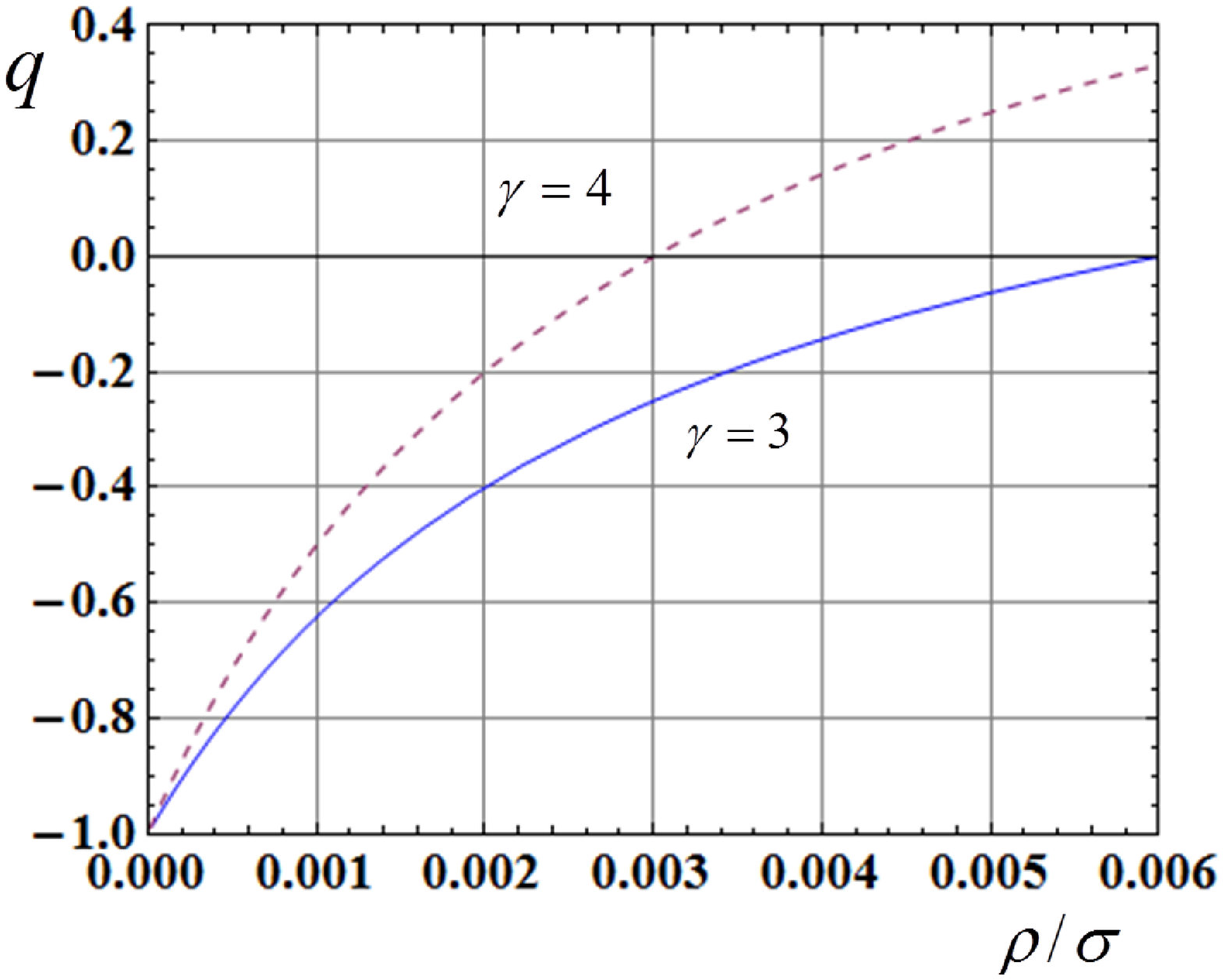,height=2.5in,width=3.35in}
\caption{As in Fig. 8 but now $\nu=10^2$ (top plot) and $\nu=10^3$
(bottom plot).}
\end{center}
\end{figure}

\medskip

In Figs.~8 and 9, we show a parametric plot between $q$
(deceleration parameter) and the ratio $\rho/\sigma$. The period
of deceleration prior to the late-epoch acceleration becomes
longer for $\nu$ larger than unity; the deceleration of the
Universe is also slower (as compared to the $\nu= 1$ case). The
Universe enters into an accelerating phase at a relatively late
time if $\nu\gg 1$. The above result reveals a genuine possibility
of realizing four-dimensional cosmology for which the Universe
decelerates between the two periods of cosmic acceleration, i.e.
between the primordial inflation and the late-epoch acceleration
at a much lower energy scale~\cite{Riess:1998,Perlmutter:1998}.


\medskip

The positivity energy condition ($\rho>0$) plus the condition
$H_0^2 \gtrsim 0$ implies that
\begin{equation}
\nu \equiv {\sigma\over 6 H_0 M^3} \gtrsim 1.\end{equation} $H_0$
may be taken to be the present Hubble scale $\bar{H}_0=2.1332
h_0\times 10^{-42}\,\text{GeV} \sim 1.5\times
10^{-42}\,\text{GeV}$ on sufficiently large scales ($h_0\simeq
0.71$ following.~\cite{Efstathiou:2013})~\footnote{This
fine-tuning may be taken as a restatement in the brane-world
scenario of the cosmological constant problem and we do not
attempt to solve it here.} Hence  \be \sigma^{1/4} \gtrsim
200.45\,\text{GeV}.\ee In fact, the condition $\nu\gtrsim 1$ also
implies \be 8\alpha H_0^2 \ll M^2,\qquad
\beta_0=4\alpha\psi_0^2\simeq 0 \ee (on sufficiently large scales)
which are obviously always satisfied [cf. Eq.~(\ref{def-H02})].
The last condition $\beta_0\simeq 0$ further implies that 
\be -\lambda M^2 \simeq {{\cal E}\over a_0^4} \ee to a large
accuracy. This result is {\it not} unnatural though -- the cosmic
expansion of our Universe could naturally take us into a state of
equilibrium where the bulk cosmological constant
$(-\Lambda\Z{5}/3)\equiv \lambda M^2 $ in five dimensions equals
the contribution of the radiation energy from the bulk. This is
also a manifestation of
AdS-gravity/Friedmann-Lama\^itre-Robertson-Walker cosmology
correspondence or AdS holography. In the limit $\lambda\to 0$, the
bulk spacetime is Minkowski flat, which means ${\cal E}=0$. The
bulk radiation term is a measure of Weyl curvature which must
vanish if $\lambda= 0$.

\medskip

A 3-brane tension of the order of $(200 \text{GeV})^4$ is in
minimum range and nucleosynthesis bounds are satisfied even for a
low value, such as $\sigma> (100 ~{\text
MeV})^4$~\cite{Copeland:2005,Ish012}. The observed cosmic
acceleration of the Universe may {\it not} be a recent phenomena,
which could have rather kicked in when $\rho \simeq \rho\Z{\text
M} + \rho_{\text R} < \left(238\,\text{GeV}\right)^4$, which means
$\rho^{1/4}$ is already $\sim 10^{13}$ times less density than the
energy scale at the end of inflation.

\medskip

Here we make one more remark. At a late epoch the effects of the
GB term (or the ${\cal R}^2$ corrections) is negligibly small; the
model is indistinguishable from the RS model except that all the
bounds found in this paper are nonexistent in RS cosmology. Of
course, the ${\cal R}^2$-type corrections are important at the
earliest epoch, whose contribution diminishes rapidly after
inflation (more precisely, after reheating) all the way to the
epochs of baryogenesis, nucleosynthesis, and at the present epoch.
This can be understood also by looking at the Lagrangian: at late
epochs (and on sufficiently large scales) $\alpha{\cal
R}^2/M^{2}\propto \alpha H_0^2 (H_0^2/M^2)\ll H\Z{0}^2$, while the
contribution of the Einstein-Hilbert term $R\propto H_0^2$.

\section{Conclusion}

The evidence of a direct detection of the primordial ``B-mode"
polarization of the CMB by BICEP2 telescope~\cite{BICEP2}, with a
relatively large tensor-to-scalar ratio $r\sim 0.19\, (+0.007 -
0.005)$ (or $r=0.16^{+0.06}_{-0.05}$ after subtracting an
estimated foreground), may be viewed as a cosmological
gravitational wave signature of primordial inflation. A large
value of $r$, along with a large value of the energy scale of
inflation, $V_*^{1/4}\sim 2\times 10^{16}~\text{GeV}$, naturally
point to some modification of Einstein gravity at a scale relevant
to inflation. In this paper, for the first time in the literature,
we identified a concrete gravitational theory where inflation has
natural beginning and natural ending. Inflation is driven by a
cosine-form potential, $V(\phi)= \Lambda^4
\left(1-\cos(\phi/f)\right)$. The effect of the ${\cal R}^2$-terms
on the magnitudes of scalar and tensor fluctuations and spectral
indices are shown to be important at the energy scale of
inflation. The model is trustworthy since a variation of the
inflaton field can be smaller than the reduced Planck mass
$M\Z{P}$. The results obtained in this paper are available also
for GB assisted $m^2\phi^2$ inflation [with $m \equiv
\Lambda^2/(\sqrt{2} f)\sim 1.0\times 10^{14}\,\text{GeV}$] as it
is a limiting case of GB--assisted natural inflation at a slightly
lower energy scale than $V_*^{1/4}$.

\medskip

The GB--assisted natural inflation is in agreement with Planck
data for a wide range of the energy scales for inflation and the
number of {\it e}--folds. The model generates a suppression in
scalar power at large scales along with reasonable amplitudes of
primordial scalar and tensor perturbations. The GB coupling
constant in the range $\alpha\sim ({10}^2-{10}^4)$ can lead to
observationally preferred values, such as, $n_s=0.9603\pm 0.005$
and $r\sim 0.14-0.21$; the latter bound is compatible with the
BICEP2 result~\cite{BICEP2}. Another important prediction of the
model is that the ratio $n\Z{T}/r$ is about (13\%-24\%) less than
the value predicted for single-field, slow-roll inflation models
based in Einstein gravity ($n\Z{T}/r= - 0.125$); the ${\cal
R}^2$-type corrections in the Lagrangian enhance the ratio of the
tensor and scalar perturbation amplitudes and hence lower the
ratio $n\Z{T}/r$. This gives a novel and testable prediction for
the GB-assisted natural inflation model.

\medskip

The model is natural and well motivated in the context of both
particle physics and high--scale string--theory models. It is
compatible with CMB data from Planck and BICEP2 experiments as
well as low red-shift data from type I supernovae. The latter
provides a direct observational evidence for an accelerating
expansion of the Universe~\cite{Riess:1998}. So, it may be the
correct description of both the early universe cosmology and
concurrent universe undergoing an extremely slow accelerating
phase in the last few billion years. For the first time in the
literature, we have presented a concrete model whose model
parameters are found in a narrow range that are consistent with
broad theoretical ideas and cosmological constraints from CMB
observations by
the BICEP2 and {Planck} telescopes. 

\medskip

A very recent paper from the {Planck}
Collaboration~\cite{Adam:2014bub} [{Planck} intermediate results.
XXX] appears to show that the BICEP2 gravitational wave result
could be due to the dust contamination. This new analysis does not
completely rule out BICEP2's original claim just yet -- detailed
cross-correlation studies of Planck and BICEP2 data would be
required for a definitive answer. Nevertheless, the results in
this paper are purely theoretical and they are natural outcomes of
a ``natural inflation model" that takes into account the
contributions of ${\cal R}^2$ terms in the Lagrangian, which is
separately well motivated. In fact, the model can still satisfy
the Planck constraint $r< 0.13$ provided that the scalar spectrum
spectral index is in high end of the $1\sigma$ result, namely,
$n_s\gtrsim 0.967$, and/or the number of {\it e}--folds
$N_*\gtrsim 60$.

\medskip

Cosmological observations when interpreted in terms of a FLRW
metric with (assumed) scale-free density perturbations imply that
$\Lambda\Z{4} \simeq 3 H_0^2$. This is then interpreted as dark
energy with $\rho\Z{\text DE} \simeq 3 M\Z{P}^2 H_0^2$. Dark
energy is a difficult problem in cosmology (see,
\cite{Copeland:2006,MLi:2011} for reviews on dark energy theory)
mainly because it requires setting the key parameter(s) to be of
order $H_0^2$ by hand. Explaining dark energy problem usually
means (i) getting the correct equation of state, (ii) getting the
right proportion of dark energy and matter (ordinary plus dark),
(iii) explaining the triple cosmic coincidence ($\rho\Z{\Lambda}
\sim \rho\Z{\text M} \sim \rho\Z{\text R}$) around the onset of
late--epoch cosmic acceleration, and finally (iv) getting the
Hubble scale that asymptotes to $H_0$ when $\rho$ gets close to
the critical density $\rho_c$. These are {\it not} independent
though -- each one of these characteristics of the ``dark energy"
problem follows simply because $H_0$ is a key physical parameter.
In this paper we have shown that instead of picking $H_0$ by hand
we can relate it with the 3-brane tension and the curvature
coupling parameters.

\medskip
\section*{ACKNOWLEDGMENTS} It is a pleasure to thank Richard Easther,
Will Kinney, Anupam Mazumdar, M Sami, Subir Sarkar, Jiro Soda and
David Wiltshire for useful discussions and helpful comments on the
draft. I am grateful to Oxford Theory and Astrophysics Groups and
Nottingham University Theory Group for their hospitality during my
visits. This work was supported by the Marsden Fund of the Royal
Society of New Zealand.

\end{document}